\documentclass[twocolumn]{aastex631}

\usepackage{bm}
\usepackage{mathtools}

\newcommand{\RLC}{R_{\rm LC}}

\newcommand{\orbinc}{i_{\rm orb}}

\newcommand{\maginc}{\alpha_{\rm B}}
\newcommand{\obsinc}{i_{\rm obs}}
\newcommand{\compactness}{u}
\newcommand{\fopen}{f_{\rm open}}
\newcommand{\Sopen}{\mathcal{S}_{\rm open}}
\newcommand{\Cminus}{C_{-}}
\newcommand{\Cplus}{C_{+}}
\shorttitle{Physics-Informed Modeling of J0740}
\shortauthors{Huang~et~al.}

\definecolor{ao(english)}{rgb}{0.0, 0.5, 0.0}

\definecolor{claudeblue}{rgb}{0.0, 0.25, 0.8}

\begin{document}

\title{Joint Mass-Radius and Magnetic-Geometry Inference of PSR J0740+6620}

\author[0000-0001-6406-1003]{Chun Huang}
\affiliation{Department of Physics and McDonnell Center for the Space Sciences, Washington University in St. Louis, St. Louis, MO 63130, USA}

\correspondingauthor{Chun Huang}
\email{chun.h@wustl.edu}

\author[0000-0001-6356-125X]{Tuomo Salmi}
\affiliation{Department of Physics, University of Helsinki, P.O. Box 64, FI-00014 University of Helsinki, Finland}

\author[0000-0002-4738-1168]{Alexander Y. Chen}
\affiliation{Department of Physics and McDonnell Center for the Space Sciences, Washington University in St. Louis, St. Louis, MO 63130, USA}
\begin{abstract}
X-ray pulse-profile modeling can measure neutron-star radii, but its physical interpretation remains limited by the largely phenomenological description of the surface hotspots. Here we connect the X-ray hotspots directly to the pulsar magnetospheric structure and jointly infer the mass, radius, and magnetic geometry of the massive millisecond pulsar PSR J0740+6620 from \emph{NICER} and \emph{XMM-Newton} observations. Using a GPU-accelerated implementation, we compare two low-dimensional magnetic configurations that can produce non-antipodal emission: a centered axisymmetric dipole--quadrupole and a shifted dipole. The shifted-dipole model is decisively favored under the adopted priors, with $\Delta\ln Z=5.65$ (Bayes factor $\sim285$), and yields
$M=2.059^{+0.068}_{-0.069}\,M_{\odot}$ and
$R=13.28^{+1.47}_{-1.07}\,\mathrm{km}$, broadly consistent with previous phenomenological analyses, while requiring nearly symmetric heating of the two polar caps. Both the marginalized shifted-dipole posterior and the timing-compatible representative require an effective X-ray-emitting footprint much smaller than the nominal force-free polar cap calibrated for a centered dipole. This compact footprint is not reproduced by our exploratory photon-conversion estimates and instead points to missing physics in the mapping from magnetospheric currents to observable X-ray emission. Our results show that physically motivated hotspots can preserve the mass--radius inference while turning hotspot morphology into a probe of pulsar magnetospheric structure and emission physics.
\end{abstract}

\section{Introduction}
\label{sec:introduction}

X-ray pulse-profile modeling (PPM) constrains neutron-star masses and radii through the rotational modulation of  surface emission. Gravitational light bending and redshift depend primarily on stellar compactness, while Doppler boosting, aberration, propagation-time delays, and rotational oblateness introduce additional radius dependence and help break the mass--radius degeneracy \citep{Pechenick_1987,Beloborodov_2002,Poutanen_2006,Morsink_2007,Watts_2016}. Energy-resolved modeling further exploits the spectrum and angular beaming set by the neutron-star atmosphere. With the long-duration, energy-resolved timing observations provided by the Neutron Star Interior Composition Explorer (\emph{NICER}), PPM has yielded mass--radius constraints for several rotation-powered millisecond pulsars when relativistic, atmospheric, instrumental, and background effects are modeled consistently \citep{Bogdanov_19,Bogdanov_21}.

Independent \emph{NICER} analyses of PSR J0030+0451 obtained the first mass--radius constraints for this source and strongly disfavored simple antipodal circular hotspots, instead favoring non-antipodal, noncircular, or multi-component emitting regions \citep{Miller2019,Riley2019}. Joint \emph{NICER} and \emph{XMM-Newton} analyses subsequently measured the radius of PSR J0740+6620 \citep{Miller2021,Riley2021}. This pulsar is particularly informative because radio Shapiro-delay timing tightly constrains its mass to $M=2.08\pm0.07\,M_{\odot}$ and orbital inclination to $i_{\rm orb}=87.56^{+0.17}_{-0.18}$ degrees \citep{Fonseca_21}. Its high mass makes the radius especially valuable for probing dense matter at high central densities, while the orbital inclination and gamma-ray observations can provide complementary information on the viewing and magnetospheric geometry \citep{Guillemot_2016}. More recent analyses using longer \emph{NICER} exposures have further refined the radius inference \citep{Dittmann24,Salmi_2024_j0740}.

Previous pulse-profile analyses of PSR J0740+6620 described the hot regions phenomenologically, allowing their positions, sizes, and component temperatures to vary independently, while typically representing the surface emission with one or more approximately isothermal regions
\citep{Miller2021,Riley2021,Dittmann24,Salmi_2024_j0740}. The inferred departure from an antipodal two-spot configuration is suggestive of an off-centered magnetic structure, although it does not uniquely imply a shifted dipole. In contrast, a physics-motivated model ties the emitting geometry and its continuous temperature distribution to an explicit magnetic-field and current-heating prescription. This tests whether a proposed magnetosphere can reproduce the observed X-ray waveform and what mass and radius it implies. A recent complementary study modeled the emitting regions using open-field-line footpoints of a static vacuum offset dipole plus axisymmetric quadrupole field while fixing several stellar and geometric parameters to values inferred from phenomenological modeling \citep{Taiyebah2026}. Here, we present the first end-to-end Bayesian pulse-profile analysis of PSR J0740+6620 in which force-free-current-informed temperature maps are fitted directly to the X-ray data while the stellar mass and radius, magnetic geometry, viewing geometry, and current-to-heating parameters are inferred jointly. This turns hotspot morphology from phenomenological freedom into a model-conditional physical observable.

The semi-analytic matched-asymptotic force-free framework of \citet{Gralla2016,Gralla2017} connects the near-surface polar-cap structure and current distribution to the magnetic field and the far-zone force-free solution. \citet{Lockhart2019} used this approach to calculate temperature maps and X-ray light curves for inclined dipolar and dipole--quadrupole configurations, while \citet{Huang:2025_hotspot} extended the hotspot calculation to non-axisymmetric, off-centered dipoles and reproduced important features of the X-ray light curves of PSR J0030+0451 and PSR J0437$-$4715. \citet{Huang2026} further derived analytic expressions for the field-aligned current invariant in inclined dipolar and general dipole--quadrupole fields, extending the efficient mapping from magnetospheric structure to polar-cap heating to a multipole-consistent current prescription.

Such physics-motivated models are computationally demanding because each sampled magnetic configuration requires evaluating a smooth surface temperature map and computing the corresponding surface emission. Although the temperature map is determined from an analytic prescription, its evaluation still requires repeated derivatives and equation solving, while the continuous temperature distribution introduces an additional dimension in the atmosphere interpolation compared with uniform-temperature hotspots. We therefore use the GPU-accelerated pulse-profile modeling framework GPU-PPM \citep{zhou2026}. Temperature-map evaluation, atmosphere interpolation, relativistic ray tracing, detector-response convolution, and multi-instrument likelihood evaluation are all performed on the GPU, making high-resolution Bayesian inference computationally practical.

In the following sections, we jointly analyze the \emph{NICER} and \emph{XMM-Newton} observations of PSR J0740+6620 to infer its mass and radius together with the magnetic geometry and current-to-heating parameters. We compare two primary magnetic configurations, a centered dipole--quadrupole and a shifted dipole, and also examine a shifted dipole--quadrupole model as an exploratory extension in the appendix. We ask which primary magnetic configuration is better supported by the data, how replacing phenomenological hotspots with a magnetospherically organized temperature distribution affects the inferred mass and radius, and whether the inferred viewing geometry, effective footprint cutoff, and relative current-to-heating normalization between the two polar caps are consistent with independent observations and existing theoretical expectations for pulsar magnetospheric and surface-heating physics.

\section{Data and Modeling Framework}
\label{sec:methods}

\subsection{Observations}

We used the processed \emph{NICER} event list described by \citet{Salmi_2024_j0740}\footnote{\url{https://doi.org/10.5281/zenodo.10519473}}.  The observations span 2018 September 21--2022 April 21 and have a total exposure of $2{,}733{,}812~\mathrm{s}$.  Events in pulse-invariant (PI) channels $30\leq {\rm PI}<150$ ($0.3$--$1.5~\mathrm{keV}$) were folded into 32 rotational-phase bins.  The resulting phase--energy array contains 1,176,389 counts in 120 detector channels (source plus background).

We fitted these data jointly with the phase-averaged \emph{XMM--Newton} EPIC-pn, MOS1, and MOS2 spectra in the same energy band. For each \emph{XMM--Newton} detector, we used the source and blank-sky spectra, ARF, and RMF provided in the same archive as the \emph{NICER} data. For both \emph{NICER} and \emph{XMM--Newton}, the response operator was RMF${}\times{}$ARF${}\times{}$input-energy-bin width. The \emph{XMM--Newton} blank-sky exposures and BACKSCAL factors entered the background constraints separately (Section~\ref{sec:likelihood}). Table~\ref{tab:xmmdata} lists the \emph{XMM--Newton} selections. We did not sample cross-calibration or effective-area factors for any instrument, so the inferred hotspot temperature and area constraints are conditional on the nominal responses and do not incorporate calibration uncertainty. 

\begin{deluxetable*}{lrrrrr}[t]
\tablecaption{XMM--Newton Data Used in the Joint Likelihood
\label{tab:xmmdata}}
\tablehead{
\colhead{Detector} &
\colhead{Channels} &
\colhead{$N_{\rm src}$} &
\colhead{$t_{\rm src}$ (s)} &
\colhead{$t_{\rm blank}$ (s)} &
\colhead{BACKSCAL}
}
\startdata
pn   & 57--298 & 94 & 6808.730  & 451098  & 0.9212 \\
MOS1 & 20--99  & 59 & 17959.574 & 1576230 & 1.074 \\
MOS2 & 20--99  & 67 & 18680.734 & 1512560 & 1.260 \\
\enddata
\tablecomments{
The pn selection spans $0.303906$--$1.50264~\mathrm{keV}$;
the MOS selections span $0.3$--$1.5~\mathrm{keV}$.
$N_{\rm src}$ is the number of counts in the selected source region.
$t_{\rm src}$ and $t_{\rm blank}$ are the source and blank-sky exposures,
and BACKSCAL is the supplied geometric scaling factor for the extraction
region.
}
\end{deluxetable*}

\subsection{Current-informed temperature template}
\label{sec:template}

Our physical hypothesis is that the thermal X-ray pattern follows the surface footprint of the magnetospheric current, rather than a freely drawn set of spots. In a stationary force-free magnetosphere, the electromagnetic field determines the four-current required to maintain it. This fixes the location and relative structure of the candidate heating regions once the magnetic geometry is specified. It does not fix the gap voltage, pair multiplicity, returning-particle fraction, or heat deposited per particle. We therefore derive the surface current distribution from the force-free prescription while fitting a small number of parameters that map this current to surface heating.

We describe the near-zone magnetic field using the Euler-potential formulation of \citet{Gralla2017}, to leading order in $\epsilon=\Omega R/c$, where $\Omega\equiv|\bm\Omega|=2\pi\nu$ is the spin angular frequency:
\begin{equation}
    \bm B=\nabla\alpha\times\nabla\beta.
\end{equation}
Here $\alpha$ labels magnetic-flux surfaces and $\beta$ distinguishes individual field lines on each surface.

To describe a displaced magnetic source, we follow the construction of \citet{Huang:2025_hotspot}. Let $\bm r$ denote the position relative to the stellar center and $\bm s=(x_s,y_s,z_s)$ the physical displacement of the magnetic source. We define the dimensionless source-centered coordinate
\begin{equation}
    \bm r'=\frac{\bm r-\bm s}{R},
\end{equation}
with $r'=|\bm r'|$. Primed spherical coordinates are centered on the magnetic source, with the polar axis aligned with the magnetic moment; $\theta'$ and $\phi'$ are therefore the magnetic colatitude and azimuth. For the aligned, axisymmetric multipoles considered here, we take $\beta=\phi'$. The sampled displacement is $\bm s/R=(x_s/R,y_s/R,z_s/R)$.

For the dipole--quadrupole configurations, we use the multipolar extension of \citet{Huang2026}, for which the flux function is
\begin{equation}
\alpha(\bm r)=
\mu\frac{\sin^2\theta'}{r'}
+q_\alpha\frac{\sin^2\theta'\cos\theta'}{r'^2},
\label{eq:alpha}
\end{equation}
where the first term is dipolar and the second is an axisymmetric quadrupolar perturbation. We set the effective dipole moment to $\mu=1$ for normalization, and $q_\alpha$ controls the quadrupolar contribution. When $\bm s\ne0$, the displacement applies to the complete near-zone field. A fully general quadrupole would introduce additional degrees of freedom and could produce different current and temperature maps; we do not explore this larger model space in order to limit the number of free parameters.

For the boundary of the open-field region, we adopt the nominal force-free value of the flux function on the last open field line \citep{Gralla2017,Lockhart2019},
\begin{equation}
\alpha_{0,\mathrm{FF}}=
\left(\frac{3}{2}\right)^{1/2}
\mu\,\Omega
\left(1+0.2\sin^2\maginc\right),
\label{eq:alpha0}
\end{equation}
where $\maginc$ is the angle between the spin and magnetic axes. The observer inclination $\obsinc$, introduced below, is instead the angle between the spin axis and the line of sight.

We set $\alpha_0\equiv\fopen\alpha_{0,\mathrm{FF}}$ and define the model open footprint by $0<\alpha<\alpha_0$. The lower bound excludes negative-$\alpha$ surfaces that do not connect to the adopted dipole-dominated far zone. Equation~(\ref{eq:alpha0}) provides a leading-order reference when the field is dipole dominated on scales approaching the light cylinder, even if higher multipoles strongly distort the surface footprints \citep{Gralla2016,Gralla2017}. Such higher multipoles can also arise from a displaced dipole. The moderate changes in spin-down found in global force-free simulations of off-centered dipoles further support adopting the dipolar open-flux scale as an approximate reference \citep{P_tri_2020}, even though the displacement can substantially reshape the surface polar caps. We have not quantified the accuracy of this reference for the geometries considered here using global magnetospheric calculations.

The nominal footprint also neglects the general-relativistic modification of the stellar field. The four-current in our model is evaluated covariantly and includes frame dragging \citep{Huang:2025_hotspot}, but Equation~(\ref{eq:alpha}) is a flat-spacetime potential, because no relativistic solution is available for a displaced dipole. For a centered dipole, the Schwarzschild exterior solution retains the $\sin^2\theta'$ dependence but enhances the surface value of the flux function by
\begin{equation}
\Delta_1(\compactness)=-\frac{3}{8\compactness^3}\left[\ln(1-2\compactness)+2\compactness+2\compactness^2\right],
\label{eq:delta1}
\end{equation}
where $\compactness\equiv GM/(Rc^2)$ is the stellar compactness \citep[$C=2\compactness$ in the notation of][]{Gralla2016,Gralla2017}. Because $\alpha_{0,\mathrm{FF}}$ is fixed by the far-zone dipole moment, the open flux is concentrated into a footprint smaller by $1/\Delta_1$ \citep[see Figure~1 of][]{Gralla2017}, i.e., by factors of $0.71$ and $0.65$ at $\compactness=0.19$ and $0.23$, respectively. For a centered dipole this is a uniform rescaling of $\alpha$ on the stellar surface and is therefore exactly degenerate with $\fopen$: the relativistic nominal footprint corresponds to $\fopen\simeq1/\Delta_1$ rather than unity.

The open flux can also depend on the plasma supply and departures from the force-free approximation. We therefore interpret $\fopen$ as an effective flux-cutoff parameter that can accommodate uncertainty in the nominal open flux, including its relativistic normalization, and incomplete heating of the physical open cap. In this sense, $\fopen$ absorbs our uncertainty in the hotspot area relative to the nominal force-free scale adopted here. The effective emitting region may be smaller, for example, if photon conversion occurs at finite altitude rather than directly near the stellar surface, or if the heating is concentrated into smaller-scale structures, such as regions associated with current-sheet heating. In the small-cap dipolar limit, the selected footprint area and angular radius scale approximately as $\fopen$ and $\fopen^{1/2}$, respectively; the actively heated area also depends on the heating prescription. We also define the open-footprint area ratio $\Sopen$ as
\begin{equation}
\begin{aligned}
\Sopen&\equiv
\frac{A_{\rm open}(\fopen)}{A_{\rm open}(1)},\\
A_{\rm open}(f)&\equiv
\int H\!\left[\alpha(\theta,\phi)\right]
H\!\left[\alpha_0(f)-\alpha(\theta,\phi)\right]\,dA ,
\end{aligned}
\label{eq:open_area_ratio}
\end{equation}
where $H(x)$ is the Heaviside step function, defined as $H=1$ for $x>0$ and $H=0$ for $x<0$. The numerator and denominator are evaluated at identical values of $\bm s/R$, $q_\alpha$, $R$, and $\maginc$. We evaluate Equation~(\ref{eq:open_area_ratio}) on a $256\times256$ surface quadrature for all equal-weight posterior samples from each model. $\Sopen$ directly measures the change in the integrated surface area satisfying $0<\alpha<\alpha_0$ relative to the nominal $\fopen=1$ footprint, and is therefore a more direct measure of the hotspot-area change.

The field-aligned current is set by the scalar $\Lambda(\alpha,\beta)$, which is constant on a magnetic field sheet at the order retained. The regular dipolar solution contains $J_0$ and $J_1$ Bessel-function terms, multiplied by $\cos\maginc$ and $\sin\maginc\cos\beta$, respectively \citep{Gralla2017}. An axisymmetric quadrupole rotated into the spin frame adds $m=0$, 1, and 2 terms, represented by $J_0$, $J_1\cos\beta$, and $J_2\cos2\beta$, with the corresponding Wigner rotation coefficients \citep{Huang2026}. The amplitude of this quadrupolar contribution to the current invariant is denoted by $q_\Lambda$. The distinction between $q_\alpha$ and $q_\Lambda$ is physical: $q_\alpha$ controls the near-zone field and hence the cap shape, whereas $q_\Lambda$ controls how the outer magnetosphere closes the quadrupolar current. For a centered field, leading-order matching gives $q_\Lambda\simeq\kappa(2/\sqrt{6})\epsilon q_\alpha$, where $\kappa$ is an order-unity geometric factor. Neither $\kappa$ nor the corresponding relation for a shifted multipole is known accurately enough to impose as a prior constraint. We therefore infer the two amplitudes independently. The separatrix return-current sheet, localized near the boundary $\alpha=\alpha_0$ between open and closed field lines, is unresolved by our smooth current prescription. We therefore exclude its surface footprint from the heating map.

At each surface element, we evaluate the charge density $\rho_e$ and spatial current $\bm J$ in a local orthonormal frame. The invariant $J^\mu J_\mu=|\bm J|^2-\rho_e^2$ (assuming $c=1$) separates spacelike currents, which cannot be supplied by a single charge species, from timelike currents. Together with the sign of $J^{\hat r}/\rho_e$, it identifies the current systems most likely to require pair creation and a returning beam \citep{Huang:2025_hotspot,Huang2026}. Force-free electrodynamics alone does not specify the kinetic transition between these regimes.

We consequently regard the binary activity mask $\chi_{\rm act}$ as part of the heating closure. In the sign convention used in our calculation, $\chi_{\rm act}=0$ outside the open cap and for elements satisfying both $J^\mu J_\mu<0$ and $J^{\hat r}/\rho_e>0$, and $\chi_{\rm act}=1$ elsewhere. With the radial direction defined as outward, $J^{\hat r}/\rho_e$ represents the radial velocity of a single-species timelike current. Thus, $J^{\hat r}/\rho_e<0$ corresponds to inward particle flow, whereas $J^{\hat r}/\rho_e>0$ corresponds to an outward sub-Goldreich--Julian current, which is expected to be inactive \citep{ChenBeloborodov2013}.

On the remaining open-field elements, the native heating temperature is
\begin{equation}
T_J=10^6~\mathrm{K}\left[|\bm J|
\max\left(1-\frac{\alpha}{\alpha_0},0\right)\right]^{1/4}.
\label{eq:heating}
\end{equation}
The fourth root converts the adopted current-dependent deposited flux to an effective temperature. The factor approaching zero at $\alpha_0$ is a proxy for the declining gap potential toward the last open field line; without it, the cap would end at an unphysical, sharply heated edge.

The temperature passed to the atmosphere calculation is
\begin{equation}
T(\theta,\phi)=10^{\ell_{\rm floor}}~\mathrm{K}
+\chi_{\rm act}10^{\ell_T}C_\pm T_J,
\label{eq:temperature}
\end{equation}
Thus, elements with $\chi_{\rm act}=0$ receive only the baseline temperature $10^{\ell_{\rm floor}}~\mathrm{K}$, rather than being assigned zero temperature. The cap label is assigned according to the sign of $\cos\theta'$, where $\theta'$ is measured relative to the magnetic axis in the magnetic-source-centered coordinate system. Surface elements with $\cos\theta'\geq0$ belong to the positive magnetic branch and are assigned $\Cplus=10^{-\ell_C/2}$, whereas those with $\cos\theta'<0$ belong to the negative magnetic branch and are assigned $\Cminus=10^{\ell_C/2}$. Thus, $\Cminus/\Cplus=10^{\ell_C}$. These labels refer to magnetic-pole identity, not to the northern and southern stellar hemispheres.

We also define the common temperature multiplier $A_T\equiv10^{\ell_T}$. Each of the four quantities introduced in Equations~(\ref{eq:alpha0}) and (\ref{eq:temperature}) corresponds to a specific limitation of the heating calculation. The normalization $\ell_T$ absorbs the unknown field amplitude, accelerating voltage, return fraction, and thermalization efficiency; the force-free current map otherwise predicts only relative temperatures. The parameter $\ell_C$ allows the two polar caps to deposit different fractions of their power at the surface. Equal efficiencies give $\ell_C=0$, but asymmetric pair supply or heat transport need not respect this expectation even when the global current closes.

The baseline temperature $10^{\ell_{\rm floor}}~\mathrm{K}$ is included as a nuisance term representing the unspecified cold-surface temperature in regions receiving negligible or no current heating. It prevents the temperature from being exactly zero wherever $T_J$ vanishes. It should not be interpreted as an independently X-ray-bright whole-surface component: its prior lies below the lower temperature limit of the neutron-star non-magnetic atmosphere (NSX) grid implemented here, $\log_{10}(T/\mathrm{K})=5.10$, and surface elements below that limit contribute no model counts. Because this baseline is added to the current-heating temperature, its fractional effect is largest for weakly heated cells near the lower boundary of the atmosphere grid and is subdominant in the bright cap interiors. Retaining it marginalizes over the uncertain cold-surface boundary condition while preventing a cool surface component from being used to fit the data. Thus, $\fopen$ accounts for uncertainty in the effective open-flux cutoff, while $\ell_T$, $\ell_C$, and $\ell_{\rm floor}$ account for uncertainties in converting the current pattern into surface temperatures.
\subsection{Models for Stellar Magnetic Field}

A pure centered dipole is too restrictive for the substantially non-antipodal emitting geometry inferred in previous analyses of PSR~J0740+6620. We therefore use two primary models that add the simplest distinct kinds of magnetic freedom. The centered dipole--quadrupole model fixes $\bm s/R=0$ and varies $q_\alpha$ and $q_\Lambda$. It tests whether a surface multipole and its associated current redistribution are sufficient to explain the observations. The shifted-dipole model varies $(x_s/R,y_s/R,z_s/R)$ and sets $q_\alpha=q_\Lambda=0$. It tests whether displacement of the dipole field alone can account for the non-antipodal emission geometry. Including the stellar, viewing, heating, and phase parameters, these models have 11 and 12 dimensions, respectively. We also calculate a 14-dimensional shifted dipole--quadrupole extension that allows all five magnetic quantities to vary. Because the present data do not require this additional freedom, its detailed results are deferred to Appendix~\ref{app:shifted_dq}.

\subsection{Relativistic Waveform and Joint Likelihood}\label{sec:likelihood}

Waveforms were calculated with GPU-PPM \citep{zhou2026} \footnote{}. We used the oblate-Schwarzschild approximation \citep{Morsink_2007,AlGendy_14}, including the latitude-dependent surface shape and normal, gravitational redshift and light bending, Doppler boosting, aberration, propagation-time delays, and the projected emission angle. The comoving specific intensity was interpolated from a fully ionized hydrogen NSX atmosphere table in effective temperature, surface gravity, photon energy, and emission angle \citep{Ho_2001}.For all instruments, interstellar absorption \citep[using the \texttt{tbnew} prescription;][]{wilms_00} and detector redistribution were included in the tabulated instrumental operator.

We fixed the hydrogen column density to $N_\mathrm{H}=2.0\times10^{20}\,\mathrm{cm}^{-2}$ in all our analyses. This value lies within the 68\% credible interval obtained by \citet{Salmi_2024_j0740} when $N_\mathrm{H}$ was sampled using the same dataset. However, fixing $N_\mathrm{H}$ means that its uncertainty and correlations with the hotspot parameters are not propagated into our results. The inferred hotspot constraints are therefore conditional on this choice. Fixing $N_\mathrm{H}$ and the effective-area normalizations removes nuisance directions that can trade soft X-ray attenuation and absolute flux against hotspot temperature and area. \citet{Salmi_2024_j0740} sampled both $N_\mathrm{H}$ and correlated instrumental scale factors and found their radius inference robust to reasonable changes in the calibration priors. That result is reassuring for the inferred stellar radius, but it does not demonstrate that the hotspot properties are unchanged in our different surface-heating model. The inferred cap contraction and temperature normalization should therefore be interpreted as conditional on the fixed absorption and nominal instrument responses adopted here.

The temperature template was constructed on a $256\times256$ grid, interpolated onto a \texttt{HEALPix} surface tessellation with resolution parameter $N_{\rm side}=40$, and ray traced from those surface elements \citep{Gorski2005}. We included propagation-time delays when mapping emission phase to observer phase. The \emph{NICER} calculation used 32 emission-phase interpolation nodes and evaluated each of the 32 observed phase bins using one-point Gauss--Legendre quadrature. Each phase-averaged EPIC spectrum used four observer-phase quadrature points and 16 emission-phase interpolation nodes. The light-bending table contained $256\times512\times512$ points in compactness, deflection, and emission angle.

Temperature-map generation, ray tracing, atmosphere interpolation, response convolution, and evaluation of the scalar joint likelihood were all performed on the GPU. Parameter vectors were passed from the CPU to the GPU, while likelihood values were returned to the CPU.

Let $\bm\vartheta$ denote the sampled model parameters. Let $d_{Xjp}$ be the observed counts for instrument $X$, detector channel $j$, and phase bin $p$, and let $\lambda^{\rm src}_{Xjp}(\bm\vartheta)$ be the corresponding predicted source counts. The joint log likelihood is
\begin{equation}
\ln\mathcal{L}(\bm\vartheta)=
\ln\mathcal{L}_{\rm NICER}+\ln\mathcal{L}_{\rm pn}
+\ln\mathcal{L}_{\rm MOS1}+\ln\mathcal{L}_{\rm MOS2}.
\label{eq:jointlike}
\end{equation}
For every detector channel, we marginalized over a phase-independent background rate $b_{Xj}$. Defining the total expected counts in each phase bin as
\begin{equation}
\lambda_{Xjp}(\bm\vartheta,b_{Xj})
\equiv \lambda^{\rm src}_{Xjp}(\bm\vartheta)
+\frac{t_X b_{Xj}}{N_{\phi,X}},
\end{equation}
the channel likelihood is
\begin{equation}
\mathcal{L}_{Xj}\propto
\int_{b^-_{Xj}}^{b^+_{Xj}}
\left[
\prod_p
\frac{\lambda_{Xjp}^{\,d_{Xjp}}
e^{-\lambda_{Xjp}}}{d_{Xjp}!}
\right]\,db_{Xj},
\label{eq:bgmarg}
\end{equation}
including the Poisson factorial terms. Here $b^-_{Xj}$ and $b^+_{Xj}$ are the lower and upper integration bounds, $t_X$ is the source exposure, and $N_{\phi,X}$ is the number of phase bins, with $\ln\mathcal{L}_X=\sum_j\ln\mathcal{L}_{Xj}$.
For each pn, MOS1, and MOS2 channel, the blank-sky counts were scaled using the source-to-blank-sky exposure ratio and the corresponding BACKSCAL factor. We then adopted an integration interval with bounds set to $\pm4$ standard deviations about the scaled blank-sky estimate, with the lower bound truncated at zero. This treatment lets the pulsed signal determine the source contribution without imposing a spectral model on the detector background. Because the \emph{NICER} backgrounds have no finite upper bound, the EPIC data also retain independent information about the phase-averaged source flux. The EPIC source spectrum was averaged over rotational phase before the response and likelihood calculation, so $N_{\phi,X}=1$ for those instruments.

The effective computation cost of the complete joint likelihood was $6.8$--$11.7~\mathrm{ms}$ per forward-model evaluation, as estimated from the sampling time.  We tested the numerical reproducibility of the predicted counts at the maximum-likelihood point from each of the three models. For each of the three maximum-likelihood parameter vectors, we recomputed the predicted counts using the independent CPU implementation. For this diagnostic, we added the channel-wise background prediction evaluated at the corresponding maximum-likelihood point to the source prediction.
The comparison was performed over the 120 PI channels and 32 phase bins used for \emph{NICER}, and over the phase-averaged 242, 80, and 80 detector channels used for pn, MOS1, and MOS2, respectively.  The maximum absolute difference in any detector bin was less than $5\times10^{-10}$ counts for \emph{NICER} and $5\times10^{-12}$ counts for EPIC.  

\subsection{Priors}\label{sec:priors}

Table~\ref{tab:priors} lists all sampled parameters, but the reasons for the adopted ranges are summarized here. Uniform distributions are denoted by $U(a,b)$. The radio Shapiro-delay measurement supplies the informative mass prior, $N(2.08,0.07^2)\,M_\odot$ \citep{Fonseca_21}; truncation at $1.6$ and $2.4\,M_\odot$ removes only the far tails while keeping the numerical domain finite. We used the broad radius interval $6$--$20~\mathrm{km}$ so that the X-ray data, rather than an assumed equation of state, determine the allowed compactness $\compactness\equiv GM/(Rc^2)$.

We adopted uniform priors on $\cos\maginc$ and $\cos\obsinc$ over $[0,1]$, corresponding to isotropic orientations within the selected angular ranges $[0,\pi/2]$. These angular restrictions are modeling assumptions: our adopted priors do not guarantee that configurations outside these ranges are represented by equivalent configurations with the same prior weight within them. Unlike previous \emph{NICER} analyses of PSR J0740+6620 that used timing-informed inclination priors under approximate spin--orbit alignment \citep{Miller2021,Riley2021,Dittmann24,Salmi_2024_j0740}, we do not concentrate the observer-inclination prior near the measured orbital inclination. This allows solutions that may require substantial spin--orbit misalignment and lets the X-ray waveform constrain the viewing geometry. The phase offset is uniform over one complete rotation because the absolute phase origin does not constrain the magnetic geometry.

The temperature-related priors were assigned in logarithmic variables because their physical normalizations are uncertain by orders of magnitude. The range $\ell_T\in[-2,2]$ permits the current-defined temperature scale to vary by a factor of $10^{-2}$--$10^2$ relative to the nominal $10^6~\mathrm{K}$ template. This deliberately broad interval encompasses uncertainty in field strength, gap voltage, pair loading, and the fraction of particle energy thermalized in the atmosphere. The floor prior, $\ell_{\rm floor}\in[2,4.5]$, corresponds to $10^2$--$3.16\times10^4~\mathrm{K}$. It keeps the unheated surface cooler than the X-ray-emitting caps and normally subdominant in the fitted band. We allowed $\ell_C\in[-0.7,1.7]$, corresponding to a negative-to-positive magnetic-branch temperature-normalization ratio $C_-/C_+$ between approximately $0.2$ and $50$. An exploratory calculation with $\ell_C\in[-0.2,0.2]$ yielded a posterior extending toward the upper boundary, motivating a wider adopted prior with greater extent toward positive values. This range includes equal normalizations and substantial asymmetry between the magnetic branches.

The nominal force-free prediction lies at $\fopen=1$ within the adopted range $\fopen\in[0.1,1.5]$. In the small-cap dipolar limit, this spans approximate cap-area ratios of $0.1$--$1.5$ and cap-radius ratios of $0.32$--$1.22$. The broad range below unity allows for the possibility that the analytic dipolar open-flux boundary overpredicts the effective heating area when only part of the nominal open cap sustains pair-producing return currents. Conversely, the upper range admits a modest enlargement if the true force-free or kinetic open flux exceeds the nominal calibration. Inferring this parameter therefore tests whether the cap area required by the data is compatible with the magnetospheric expectation; an extreme inferred value is physically informative even if the waveform can be fitted.

For shifted models, $|\bm s|/R$ is uniform in volume inside a sphere of radius $0.95$, rather than uniform independently in its Cartesian components. This choice assigns equal prior density per unit stellar volume and prevents the magnetic singularity from reaching the surface. Both quadrupole priors include zero and both polarities. The broad near-zone range $q_\alpha\in[-5,5]$ admits locally important surface-field distortions, whereas the narrower current-invariant range $q_\Lambda\in[-0.5,0.5]$ reflects the more rapid radial decay of a quadrupole and the expectation that the outer magnetosphere remains predominantly dipolar \citep{Huang2026}. It nevertheless contains the nominal centered-field mapping by a wide margin, allowing the posterior to reveal suppression, enhancement, or reversal of the quadrupolar current contribution.

The spin frequency was fixed at $\nu=346.5319965~\mathrm{Hz}$. We implemented the distance uncertainty by marginalizing over the radio-timing distance distribution inferred by \citet{Fonseca2021}, who obtained $D=1.14^{+0.17}_{-0.15}~\mathrm{kpc}$ (68.3\% credible interval). Specifically, we adopted the analytic approximation to this distance distribution used by \citet{Riley2021}, represented by a skew-normal distribution
\begin{equation}
p(D)=\frac{2}{\omega}\,
\phi\!\left(\frac{D-\xi}{\omega}\right)
\Phi\!\left[a\left(\frac{D-\xi}{\omega}\right)\right],
\end{equation}
where $\phi$ and $\Phi$ denote the probability density function and cumulative distribution function of the standard normal distribution, respectively, with shape parameter $a=1.7$, location $\xi=1.002~\mathrm{kpc}$, and scale $\omega=0.227~\mathrm{kpc}$, truncated to $0<D<1.7~\mathrm{kpc}$. A free phase offset accounts for the arbitrary phase origin of the X-ray profile.

\begin{deluxetable*}{llll}
\tabletypesize{\scriptsize}
\tablecaption{Sampled Parameters and Priors\label{tab:priors}}
\tablehead{
\colhead{Parameter} & \colhead{Meaning} & \colhead{Prior} & \colhead{Models}}
\startdata
$M$ & Gravitational mass & $N(2.08,0.07^2)\,M_\odot$, truncated to $[1.6,2.4]\,M_\odot$ & All \\
$R$ & Equatorial circumferential radius & $U(6,20)~\mathrm{km}$ & All \\
$\cos\maginc$ & Spin--magnetic-axis orientation & $U(0,1)$ & All \\
$\cos\obsinc$ & Spin--observer orientation & $U(0,1)$ & All \\
$\ell_{\rm floor}$ & $\log_{10}$ whole-surface temperature (K) & $U(2,4.5)$ & All \\
$\ell_T$ & $\log_{10}$ current-temperature scale & $U(-2,2)$ & All \\
$\ell_C$ & $\log_{10}$ relative cap normalization & $U(-0.7,1.7)$ & All \\
$\fopen$ & Open-flux rescaling & $U(0.1,1.5)$ & All \\
$\phi_0$ & Rotational phase offset (cycles) & $U(0,1)$ & All \\
$(x_s/R,y_s/R,z_s/R)$ & Dimensionless magnetic-source displacement & Uniform volume, $|\bm s|/R\leq0.95$ & Shifted models \\
$q_\alpha$ & Quadrupole amplitude in $\alpha$ & $U(-5,5)$ & Quadrupolar models \\
$q_\Lambda$ & Quadrupole amplitude in $\Lambda$ & $U(-0.5,0.5)$ & Quadrupolar models \\
\enddata
\tablecomments{``All'' denotes all calculated magnetic families. ``Shifted models'' denotes the shifted dipole and shifted dipole--quadrupole families; ``quadrupolar models'' denotes the centered and shifted dipole--quadrupole families. Fixed quantities, detector backgrounds and distance marginalized within the likelihood.}
\end{deluxetable*}

\subsection{Nested Sampling}

Posterior sampling and evidence calculation used the reactive nested-sampling implementation in \textsc{UltraNest} \citep{Buchner2021}. Nested sampling was chosen because the phase, viewing geometry, magnetic displacement, and quadrupolar parameters produce curved degeneracies and potentially separated modes, while the model comparison also requires a direct calculation of the Bayesian evidence. New points within each likelihood constraint were generated with a slice sampler using mixed random directions and 96 slice steps. The trajectory length is substantially larger than twice the 11--14 sampled dimensions and was chosen conservatively to reduce the dependence of each replacement point on its initial live point and to improve mixing within anisotropic likelihood-constrained regions. Two MPI workers, each assigned to one GPU, evaluated likelihoods concurrently; this parallelization changes the throughput but not the statistical stopping rules.

We required a minimum of 6000 live points globally and 80 live points in every identified cluster. The large global population was used to reduce the probability of losing narrow magnetic-geometry modes during prior-volume compression, while the cluster requirement prevents a mode with small prior mass from being represented by only a few points. A finite fraction of the initial prior volume may lie outside the numerical domain of the forward model. For example, the compactness may fall outside the tabulated ray-tracing range, or the surface gravity may fall outside the NSX atmosphere grid. These points were assigned a constant very low likelihood. \textsc{UltraNest} detected this plateau and reactively enlarged the initial root populations from 6000 to 14,289, 13,077, and 26,833 points for the centered dipole--quadrupole, shifted-dipole, and shifted dipole--quadrupole runs, respectively. This enlargement occurred before compression of the physical likelihood contours and preserved the requested live-point resolution after the invalid region was discarded.

The reactive stopping targets were $d\ln Z=0.1$, $dD_{\rm KL}=0.5$, a remaining-evidence fraction of $0.01$, and at least 1500 effective posterior samples; we additionally set the likelihood-range tolerance to $L_\epsilon=10^{-3}$. These conditions constrain complementary aspects of the calculation. The first two bound, respectively, the bootstrap dispersion of the evidence integral and the bootstrap uncertainty of the posterior representation, whereas the remaining-evidence criterion requires the as-yet-unintegrated contribution to be no more than 1\% of the total. The $L_\epsilon$ condition permits termination if the likelihoods of the surviving points have become numerically indistinguishable, and the effective-sample requirement protects marginalized intervals from being determined by a small number of highly weighted points. Thus, convergence was not identified with a single change in $\ln Z$. A limit of $9\times10^7$ likelihood calls was imposed only as a safeguard and was not reached: the centered dipole--quadrupole, shifted-dipole, and shifted dipole--quadrupole runs used $3.75\times10^7$, $3.83\times10^7$, and $6.83\times10^7$ evaluations, respectively, over 101,303--157,056 nested-sampling iterations.

We assessed convergence from the saved nested-sampling records rather than from the termination flag alone. The usual between-chain statistic $\widehat R$ is not applicable to this calculation. The relevant checks concern the fidelity of constrained-prior sampling, the stability of the weighted posterior, and completion of the evidence integral. All three runs passed \textsc{UltraNest}'s Mann--Whitney--Wilcoxon insertion-order test, which compares the ranks of replacement points with those expected from unbiased sampling within each likelihood constraint. Their final effective sample sizes were 27,320, 22,035, and 32,589, exceeding the requested minimum by more than an order of magnitude, and the final bootstrap estimates of $dD_{\rm KL}$ were $0.45$, $0.46$, and $0.46$, below the adopted target of $0.5$. The central evidence estimates from the single integration and from 30 bootstrap realizations differed by less than 0.005 in $\ln Z$, while the fractional contribution assigned to the unresolved evidence tail was $0.010$ in each run. Over the final 10\% of the stored nested-sampling sequence, the cumulative $\ln Z$ changed by only 0.080, 0.067, and 0.067 for the three models in the same order. We also inspected the weighted posterior projections and likelihood--prior-volume traces for late-emerging modes or abrupt changes in the marginal distributions.

These diagnostics probe complementary failure modes and support convergence within the explored posterior regions. They do not replace independent repeat runs or sampler-setting comparisons and therefore cannot exclude a missed disconnected mode or establish the detailed relative weights of weak tails and separated modes. As with any finite nested-sampling calculation, they also cannot exclude an arbitrarily small, completely disconnected mode. In the present application, that risk is reduced by the large root populations, the explicit per-cluster live-point floor, the reactive enlargement of the live-point population, and the successful insertion-rank tests. The final bootstrap-inclusive uncertainties in $\ln Z$ were 0.132, 0.118, and 0.080 for the centered dipole--quadrupole, shifted-dipole, and shifted dipole--quadrupole models, respectively. These estimated uncertainties were propagated when calculating uncertainties in the log-evidence differences.
\section{Results}
\label{sec:results}

We first present the centered dipole--quadrupole model and then ask whether shifting the dipole provides a more natural description of the same data. The centered model supplies non-antipodal freedom through a quadrupole, whereas the shifted model places the asymmetry directly in the dipole geometry. Unless otherwise stated, quoted parameter constraints are posterior medians with 16th and 84th percentiles. For the representative surface maps and two-dimensional count comparisons, we show the global maximum-likelihood configuration for the centered model. For the preferred shifted-dipole model, we instead show the highest-likelihood stored sample with $\obsinc$ in the 68\% orbital-inclination interval measured by \citet{Fonseca_21}, mapped to the observer inclination under the assumption of spin--orbit alignment. This sample lies only $\Delta\ln\mathcal L=0.974$ below the global maximum. Although the global maximum remains the correct point for reporting $\ln\mathcal L_{\max}$ and the Bayesian evidence, its parameter vector is not representative of the marginalized posterior. The observer inclination, radius, and compactness all lie away from the main posterior support.  We therefore refer to it hereafter as the ``timing-compatible representative'', distinguishing it from the formal maximum-likelihood point. We use this representative only where a single parameter vector is required, the marginalized posterior, maximum likelihood, and Bayesian evidence remain unchanged. Table~\ref{tab:maxlike_parameters} lists the two global maxima and the shifted-dipole representative. By contrast, the contour bands of the marginalized light curve and energy spectrum show posterior medians and central 68\% and 95\% intervals from all posterior samples. Below, we state explicitly whether a result refers to this representative or to the marginalized posterior.

Table~\ref{tab:model_evidence} gives the maximum likelihood and Bayesian evidence results for the two primary models. The shifted dipole has a maximum log likelihood larger by 2.57 and a log evidence larger by $5.654\pm0.177$ for the adopted priors, corresponding to a Bayes factor of $\sim285$. This indicates a decisive statistical preference under the adopted priors for the shifted-dipole model over the centered model. Beyond this statistical preference, the shifted model yields a mass--radius posterior consistent with previous analyses \citep{Miller2021,Dittmann24,Riley2021,Salmi_2024_j0740} and nearly symmetric heating between the two magnetic branches. The centered model reaches its fit through a large radius and strongly asymmetric cap heating, whereas the shifted dipole does not require either compensation.

\begin{deluxetable*}{lccc}[htbp]
\tabletypesize{\scriptsize}
\tablecaption{Global Maximum-Likelihood Configurations and the Timing-Compatible Shifted-Dipole Representative}\label{tab:maxlike_parameters}
\tablehead{\colhead{Parameter or quantity} & \colhead{Centered dipole--quadrupole} & \colhead{Shifted dipole} & \colhead{Shifted representative}}
\startdata
\cutinhead{Sampled parameters}
$x_s/R$ & \nodata & $-0.06652$ & $0.07198$ \\
$y_s/R$ & \nodata & $0.2039$ & $-0.1937$ \\
$z_s/R$ & \nodata & $-0.4553$ & $-0.5276$ \\
$q_\alpha$ & $1.298$ & \nodata & \nodata \\
$q_\Lambda$ & $-0.1200$ & \nodata & \nodata \\
$M\ (M_\odot)$ & $2.270$ & $2.028$ &$2.025$ \\
$R\ (\mathrm{km})$ & $19.52$ & $15.93$ & $13.26$ \\
$\ell_{\rm floor}$ & $3.422$ & $2.240$ & $3.974$ \\
$\cos\maginc$ & $0.9190$ & $0.06719$ & $0.04720$ \\
$\cos\obsinc$ & $0.7768$ & $0.9518$ & $0.04354$ \\
$\ell_T$ & $-0.07778$ & $0.4144$ & $0.3446$ \\
$\ell_C$ & $0.7917$ & $-0.04545$ & $0.03603$ \\
$\fopen$ & $1.426$ & $0.5607$ & $0.1942$\\
$\phi_0$ (cycles) & $0.9391$ & $0.7071$ & $0.2737$ \\
\cutinhead{Derived quantities}
$\compactness$ & $0.1717$ & $0.1879$ & {$0.2255$} \\
$\maginc$ (deg) & $23.22$ & $86.15$ & {$87.29$} \\
$\obsinc$ (deg) & $39.03$ & $17.86$ & {$87.50$} \\
$\Cminus/\Cplus$ & $6.190$ & $0.9006$ & {$1.087$} \\
$\Sopen$ & $1.168$ & $0.5314$ & {$0.1726$} \\
$\sqrt{\Sopen}$ & $1.081$ & $0.7290$ & {$0.4155$} \\
$T_{\max}\ (10^6\,\mathrm{K})$ & $1.171$ & $1.884$ & {$1.414$} \\
$\langle T\rangle_{\rm X}\ (10^6\,\mathrm{K})$ & $0.8893$ & $0.8939$ & {$0.7758$} \\
\enddata
\tablecomments{The first two data columns list the global maximum-likelihood vectors. The final column lists the timing-compatible shifted-dipole representative used in the pointwise figures, selected as the highest-likelihood stored sample in the \citet{Fonseca_21} 68\% orbital-inclination interval under spin--orbit alignment. Values are rounded to four significant figures. A blank entry denotes a parameter absent from that magnetic family. Fixed inputs such as $D$, $N_{\rm H}$, responses, and stellar spin are not repeated. Detector backgrounds are not components of these sampled vectors; the two-dimensional count figures add the channel-wise background that maximizes the likelihood conditional on the listed sample.}
\end{deluxetable*}
\begin{deluxetable*}{lrrrr}
\tablecaption{Likelihoods and Bayesian Evidences\label{tab:model_evidence}}
\tablehead{
\colhead{Magnetic model} & \colhead{$k$} & \colhead{$\ln\mathcal{L}_{\max}$} &
\colhead{$\ln Z$} & \colhead{$\Delta\ln Z$}}
\startdata
Centered dipole--quadrupole & 11 & $-21864.685$ & $-21896.392\pm0.132$ & $-5.654\pm0.177$ \\
Shifted dipole & 12 & $-21862.110$ & $-21890.738\pm0.118$ & $0$ \\
\enddata
\tablecomments{$k$ is the number of sampled parameters.  Evidence differences are relative to the shifted-dipole model.  The quoted uncertainties include the bootstrap integration uncertainty described in Section~\ref{sec:methods}.}
\end{deluxetable*}

\subsection{Centered Dipole plus Quadrupole}
\label{subsec:results_centered_dq}

\begin{figure*}
\centering
\includegraphics[width=0.98\textwidth]{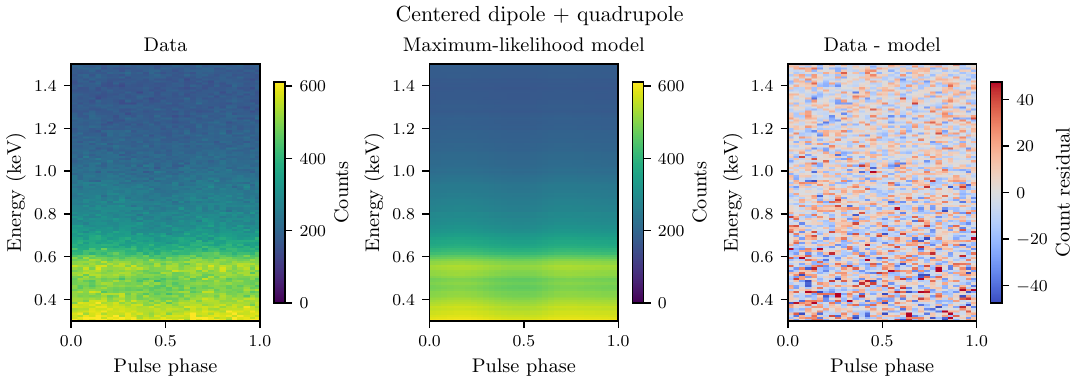}
\caption{Observed \emph{NICER} phase--energy counts (left), the maximum-likelihood centered dipole--quadrupole prediction including the channel-wise background that maximizes the likelihood for the given sample (center), and their difference (right). The parameter vector is listed in Table~\ref{tab:maxlike_parameters}.\label{fig:centered_counts}}
\end{figure*}

\begin{figure*}[t]
\centering
\includegraphics[
    width=0.92\textwidth,
    height=0.72\textheight,
    keepaspectratio
]{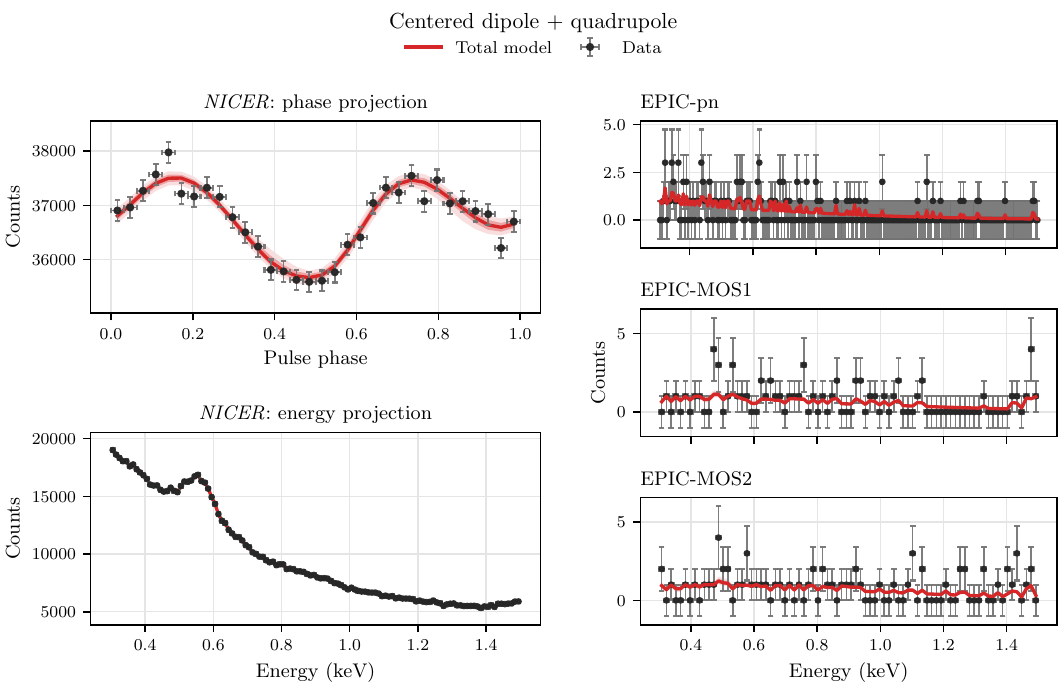}
\caption{Observed-count projections for the centered dipole--quadrupole fit. The left column shows the \emph{NICER} phase and energy projections; the right column shows the phase-averaged EPIC-pn, MOS1, and MOS2 spectra. Points have Poisson uncertainties, the solid curve is the posterior median total prediction, and the dark and light bands contain 68\% and 95\% of predictions from 4096 posterior samples.\label{fig:centered_instrument_projections}}
\end{figure*}

Because a pure centered dipole cannot produce the substantially non-antipodal geometry indicated by the data, adding an axisymmetric quadrupole is the minimal extension within a centered magnetic configuration. The influence of a centered quadrupole on the heating pattern and pulse-profile observables was discussed by \citet{Lockhart2019} and further developed in the framework of \citet{Huang2026}. This model illustrates why fitting the data is not, by itself, sufficient physical justification for a hotspot model. The maximum-likelihood phase--energy comparison in Figure~\ref{fig:centered_counts} and the posterior projection bands for \emph{NICER} and all three EPIC detectors in Figure~\ref{fig:centered_instrument_projections} show that the model can largely reproduce the main features of the data despite its lower likelihood and evidence. Nevertheless, its maximum likelihood is lower by 2.57 and its evidence is lower by $5.654\pm0.177$ in $\ln Z$ relative to the shifted dipole.

The main physical differences appear in the inferred parameter posteriors. The full parameter posterior in Figure~\ref{fig:centered_corner} shows that the centered dipole--quadrupole model favors $q_\alpha=1.22^{+0.13}_{-0.81}$ and $q_\Lambda=-0.139^{+0.039}_{-0.037}$ in its main mode, using the quadrupole to redistribute both the surface flux and the polar-cap current. The model also moves to much lower compactness, with $R=17.46^{+1.57}_{-1.48}\,\mathrm{km}$ and a median $\compactness=0.177$. The posterior probability that the radius exceeds $16\,\mathrm{km}$ is 83.8\%. The maximum-likelihood point is still more extreme, at $M=2.270\,M_\odot$ and $R=19.52\,\mathrm{km}$, near the high-radius prior boundary and in the upper tail of the radio mass prior. These values are well above the PSR~J0740+6620 radii inferred in the analyses cited above. The posterior is truncated by the $20\,\mathrm{km}$ radius prior, suggesting that the centered dipole--quadrupole hypothesis compensates for its restricted surface geometry by increasing the radius. The same model favors $\obsinc=39.85^{+9.51}_{-6.34}$ degrees, far from the nearly edge-on orbital inclination; the geometric interpretation of this result is discussed in Section~\ref{subsec:discussion_inclination}.

The associated temperature map is qualitatively very diffuse. A broad heating region covers a substantial part of the southern hemisphere, and the northern hotspot is also large. For this maximum-likelihood configuration, the local maximum temperature is $T_{\rm max}=1.171\times10^6~\mathrm{K}$, and the surface-area-weighted mean over the X-ray-active cells is $\langle T\rangle=8.893\times10^5~\mathrm{K}$. Moreover, the fit requires $\Cminus/\Cplus=6.36^{+1.65}_{-4.19}$, with a maximum-likelihood value of 6.19. Only 7.9\% of the posterior permits the cap normalizations to agree within a factor of two. Thus, the centered quadrupolar model transfers part of the geometrical burden into a strong asymmetry in the current-to-temperature conversion between the two magnetic branches.

The hotspot area is also enlarged, with $\fopen=1.275^{+0.165}_{-0.914}$ and a maximum-likelihood value of 1.426. Because the quadrupolar mapping makes the area response nonlinear, the actual footprint enlargement is milder: $\Sopen=1.111^{+0.068}_{-0.164}$, about 1.1 times the nominal force-free value. The median open region covers 24.4\% of the stellar surface. Thus, a centered dipole plus an axisymmetric quadrupole combines a broad hotspot region and a large radius with strongly asymmetric current heating to reproduce the observed pulse profile. The lower evidence shows that this compensation occupies less useful prior volume than the shifted-dipole solution. The large radius and strong magnetic-branch heating asymmetry also require more extreme physical conditions than the shifted-dipole model.

\begin{figure*}
\centering
\includegraphics[width=0.78\textwidth]{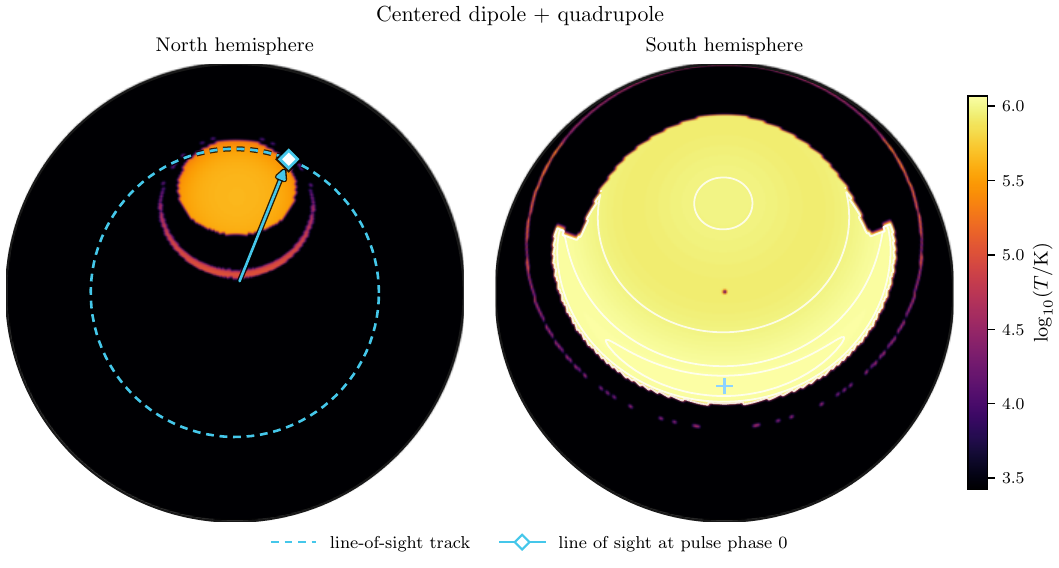}
\caption{Maximum-likelihood centered dipole--quadrupole temperature map. No longitude roll is applied to the surface view. The centered solution favors a broad heated footprint and a large contrast between the two current systems. Its local maximum temperature is $1.171\times10^6$ K, X-ray active cells weighted mean temperature is $8.893\times10^5$ K. The parameter vector is listed in Table~\ref{tab:maxlike_parameters}.\label{fig:centered_temperature_map}}
\end{figure*}

\begin{figure*}[t]
\centering
\includegraphics[
    width=0.90\textwidth,
    height=0.76\textheight,
    keepaspectratio
]{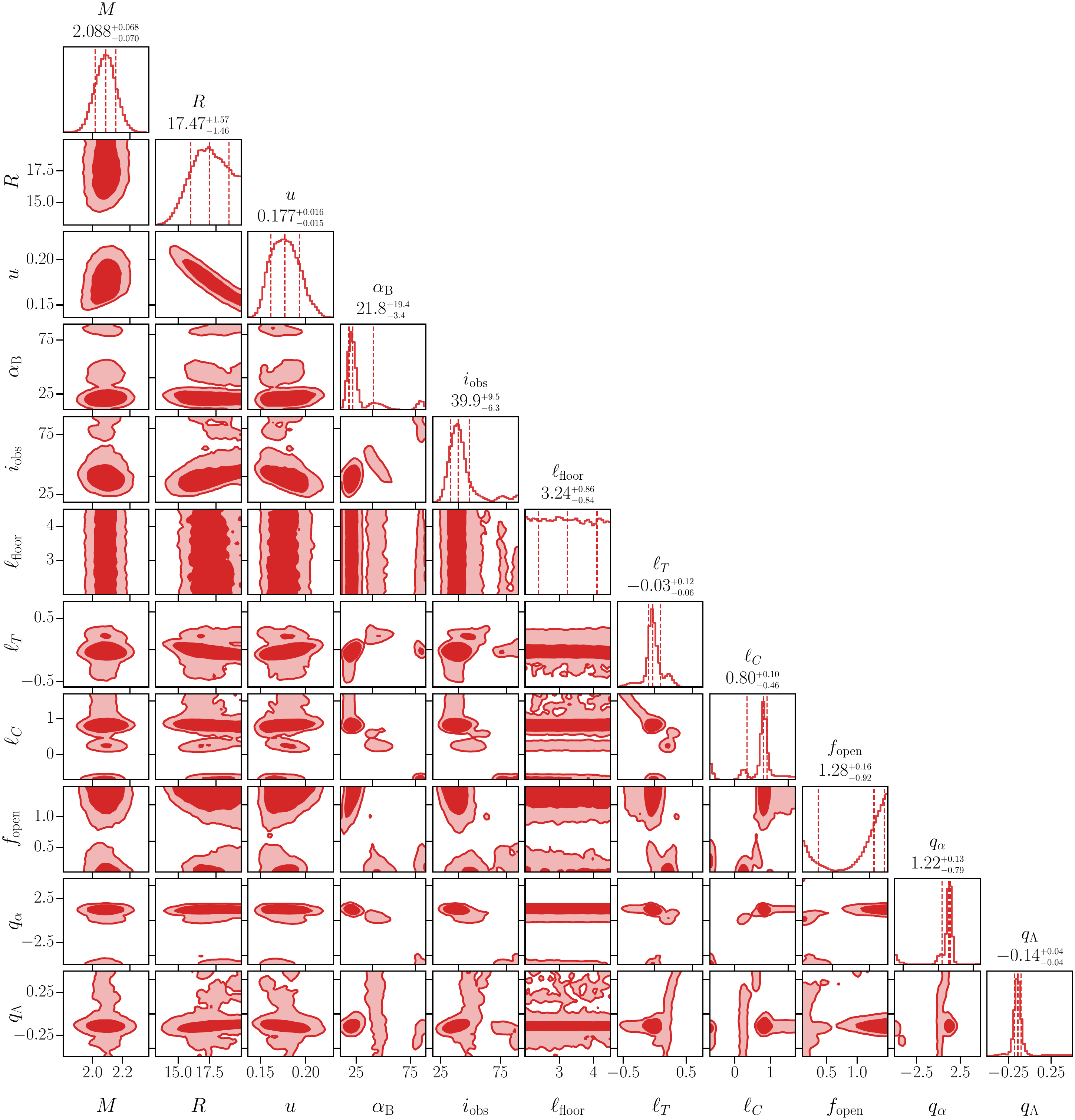}
\caption{One- and two-dimensional posterior distributions for the selected centered dipole--quadrupole model parameters, where $\compactness=GM/(Rc^2)$ is compactness.  Titles give the median and 16th--84th percentiles; contours enclose 68\% and 95\% of the smoothed posterior density. The concentration at large radius is correlated with the viewing and heating parameters and is not imposed by the radio mass prior. \label{fig:centered_corner}}
\end{figure*}

\subsection{Shifted Dipole}
\label{subsec:results_shifted}

The centered dipole--quadrupole and shifted-dipole models test two distinct mechanisms for producing non-antipodal heating. The former keeps the magnetic source at the stellar center and adds the lowest-order axisymmetric multipolar correction, whereas the latter retains a dipolar field but shifts its source from the stellar center. They are therefore not strictly ordered by complexity. The centered model has two model-specific magnetic parameters, $q_\alpha$ and $q_\Lambda$, while the shifted model has the three displacement coordinates $x_s/R$, $y_s/R$, and $z_s/R$. Previous phenomenological pulse-profile analyses of PSR~J0740+6620 have favored non-antipodal circular hotspots, a configuration suggestive of an off-centered dipole \citep{Miller2021,Riley2021,Salmi_2024_j0740,Dittmann24}. However, in those models the hot-region locations, sizes, and temperatures are not constrained by a magnetic field or magnetospheric current solution. A shifted dipole is therefore the minimal displacement-based alternative, testing whether the observed pulse profile can arise from a force-free current and heating prescription rather than from freely placed phenomenological spots.

\begin{figure*}
\centering
\includegraphics[width=0.98\textwidth]{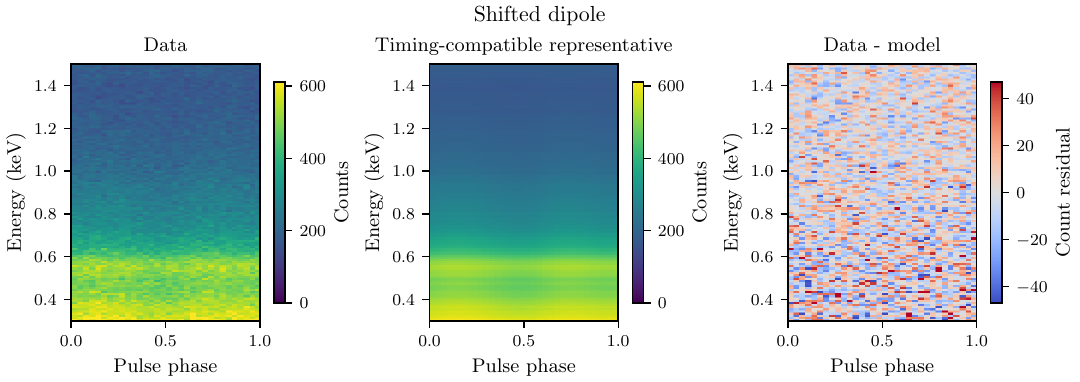}
\caption{Observed \emph{NICER} phase--energy counts (left), the timing-compatible representative shifted-dipole model including the channel-wise background that maximizes the likelihood conditional on this sample (center), and their difference (right). The common color scale in the first two panels makes the broad two-pulse morphology directly comparable. The representative solution parameter vector is listed in the final column of Table~\ref{tab:maxlike_parameters}.\label{fig:shifted_counts}}
\end{figure*}

\begin{figure*}[t]
\centering
\includegraphics[
    width=0.92\textwidth,
    height=0.72\textheight,
    keepaspectratio
]{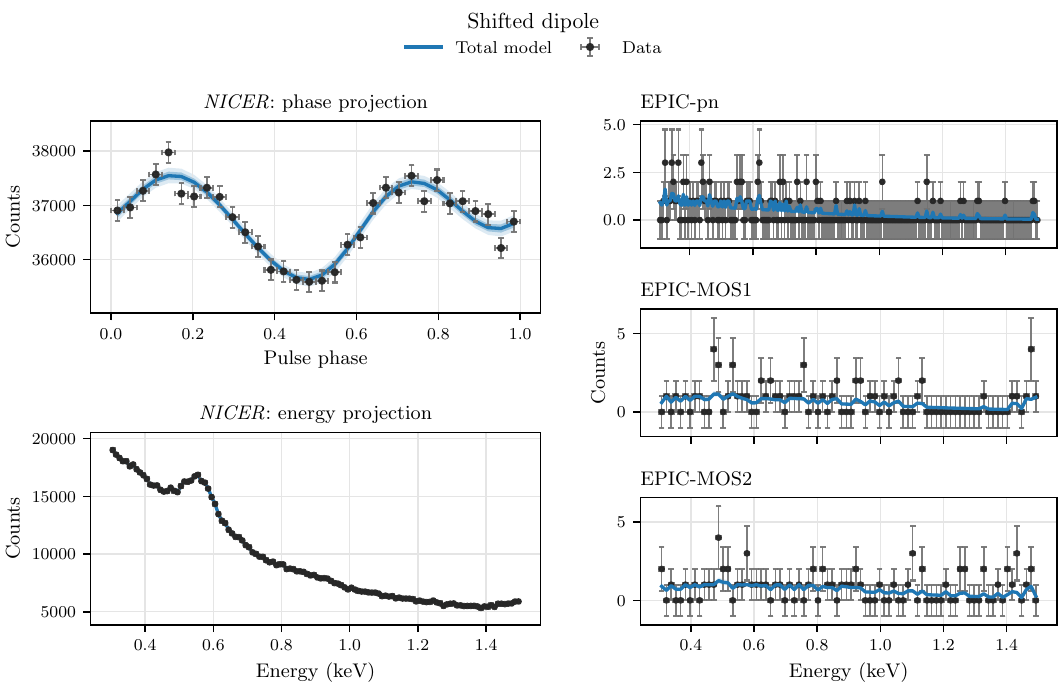}
\caption{Observed-count projections for the shifted-dipole fit. The organization and graphical elements follow Figure~\ref{fig:centered_instrument_projections}.\label{fig:shifted_instrument_projections}}
\end{figure*}

Figure~\ref{fig:shifted_counts} compares the phase- and energy-resolved \emph{NICER} counts with the timing-compatible representative prediction defined above. The model reproduces the observed pulse profile without evident systematic structure in the residual panel. Figure~\ref{fig:shifted_instrument_projections} collects the corresponding one-dimensional \emph{NICER} phase and energy projections together with the phase-averaged \emph{XMM-Newton} EPIC-pn, MOS1, and MOS2 spectra; the posterior bands show similarly good agreement for all four detectors.

Figure~\ref{fig:shifted_corner} shows the posterior of selected model parameters. We infer
\begin{align*}
M &= 2.059^{+0.068}_{-0.069}\,M_\odot, \\
R &= 13.28^{+1.47}_{-1.07}\,\mathrm{km}.
\end{align*}
The radius inferred in this work largely agrees with previous analyses based on phenomenological emitting regions \citep{Miller2021,Riley2021,Dittmann24,Salmi_2024_j0740,Hoogkamer2025}.
The radius posterior has an extended upper tail and is therefore not well represented by a Gaussian: 5.9\% of its probability lies above $16\,\mathrm{km}$ and 1.5\% above $18\,\mathrm{km}$.
A similar high-radius tail is present in the analysis of \citet{Miller2021} and in the results of \citet{Dittmann24}.
In \citet{Riley2021,Salmi_2024_j0740,Hoogkamer2025}, the tail was less pronounced, but their radius priors had an upper bound of $16\,\mathrm{km}$.
Thus, despite replacing phenomenological hotspots with a physically motivated shifted-dipole heating model, the inferred radius distribution remains broadly similar to those obtained with phenomenological surface-emission models. The mass--radius posterior remains broadly consistent when a quadrupole is added to the shifted dipole in Appendix~\ref{app:shifted_dq}, although that extension broadens the upper-radius tail. This stability indicates that the mass--radius inference is not strongly sensitive to this additional magnetic freedom.

The preferred viewing geometry is close to orthogonal, with median $\maginc\simeq80^\circ$ and $\obsinc\simeq70^\circ$, but the posterior distributions of both angles are concentrated toward the upper limits of their prior ranges. Unlike the centered model, the shifted-dipole posterior retains appreciable support near the timing-informed inclination, as quantified in Section~\ref{subsec:discussion_inclination}. The dimensionless shift coordinates $(x_s/R,y_s/R,z_s/R)$ are correlated and exhibit multimodal behavior; in particular, $y_s/R$ contains two peaks nearly symmetric about zero. Their joint posterior nevertheless supplies the non-antipodal surface geometry that the centered model lacks. The $x_s/R$ posterior is close to zero, the two modes of $y_s/R$ exclude zero, and $z_s/R$ is centered at a comparatively large nonzero displacement while still containing zero within its broad distribution. The floor-temperature parameter remains unconstrained by the data.

\begin{figure*}
\centering
\includegraphics[width=0.78\textwidth]{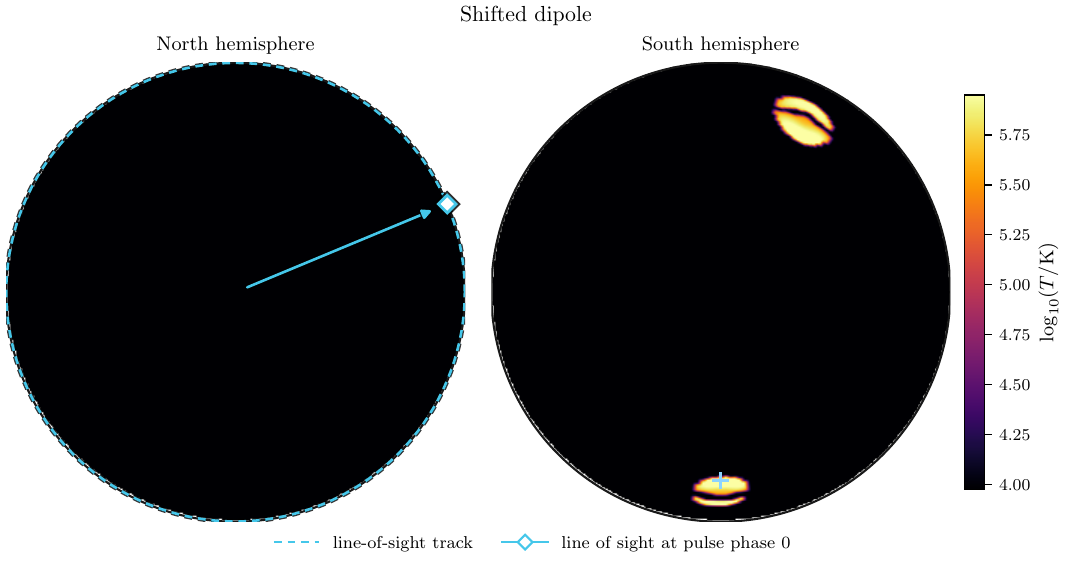}
\caption{Timing-compatible representative shifted-dipole temperature map in fixed north- and south-hemisphere orthographic views. The dashed cyan curve traces the line of sight over one rotation, while the cyan segment and diamond mark the line of sight at phase zero. The local maximum temperature is $1.414\times10^6$ K, and the X-ray-active-area-weighted mean is $7.758\times10^5$ K. The parameter vector is the final column of Table~\ref{tab:maxlike_parameters}.\label{fig:shifted_temperature_map}}
\end{figure*}

\begin{figure*}
\centering
\includegraphics[width=0.98\textwidth]{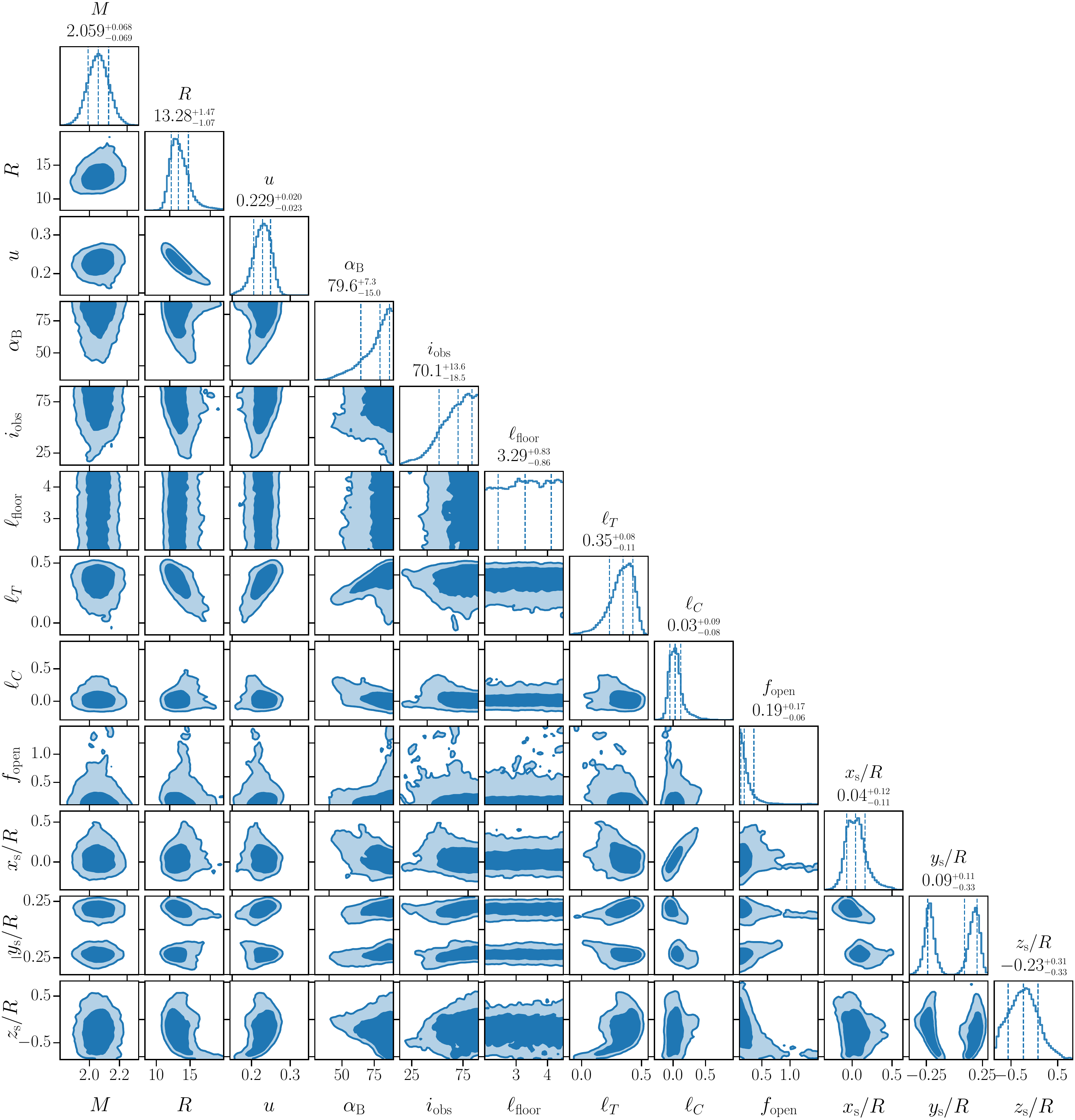}
\caption{Selected parameter posterior distributions for the shifted dipole model. Contours and titles follow Figure~\ref{fig:centered_corner}. The magnetic displacement is in units of the stellar radius.\label{fig:shifted_corner}}
\end{figure*}

The timing-compatible representative surface map in Figure~\ref{fig:shifted_temperature_map} contains two compact heated regions, with a local maximum temperature $T_{\rm max}=1.414\times10^6~\mathrm{K}$ and a surface-area-weighted mean over the X-ray-active cells of $\langle T\rangle=7.758\times10^5~\mathrm{K}$. Their interiors have comparatively broad temperature distributions, with most of the spatial variation confined to the hotspot boundaries and some unheated regions within the polar-cap footprint. This explains why uniform-temperature cap models can recover the pulse-profile data and provide similar mass--radius measurements. At the present observational precision, the data respond primarily to the location, projected area, and mean spectrum of each bright region. The physical model adds information that a uniform cap without a magnetic-field connection cannot provide. It asks whether that projected area follows from the open magnetic flux and whether the two regions can be heated with a common efficiency.

For the shifted dipole, the inferred cap heating ratio is $\Cminus/\Cplus=1.074^{+0.233}_{-0.183}$, consistent with equal current-to-heat conversion between the negative and positive magnetic branches. The timing-compatible representative shown in Figure~\ref{fig:shifted_temperature_map} has a contracted effective footprint, with $\fopen=0.1942$, $\Sopen=0.1726$, and $\sqrt{\Sopen}=0.4155$. The marginalized posterior gives $\fopen=0.187^{+0.174}_{-0.062}$, with $P(\fopen<0.5)=90.9\%$; direct surface integration gives $\Sopen=0.176^{+0.150}_{-0.059}$ and $\sqrt{\Sopen}=0.420^{+0.151}_{-0.077}$. The representative was selected using only the independently measured inclination interval and the X-ray likelihood, so the agreement in footprint was not imposed. Both the representative and the marginalized posterior require an emitting footprint smaller than the nominal $\fopen=1$ scale within the shifted-dipole model. The resulting tension between the inferred X-ray-emitting area and the nominal force-free polar-cap scale is discussed in Section~\ref{subsec:discussion_cap}.

The spectrum constrains the common current-temperature normalization to $\ell_T=\log_{10}A_T=0.346^{+0.082}_{-0.114}$. Its principal posterior dependence is the expected compensation with footprint area. The Spearman correlation coefficient between $\log\Sopen$ and $\ell_T$ is $-0.40$. For samples with $0.10<\fopen<0.15$, the median normalization is $A_T=2.43$, compared with 1.67 for $0.5<\fopen<1$. Because $A_T$ rescales the temperature map and absorbs unresolved heating physics, it is not itself a physical hotspot temperature, and we do not interpret its absolute value, or $A_T^4$, as a return-particle efficiency. The relative trend is the expected spectral response to concentrating the current-heated flux into a smaller area.

\subsection{Physical Comparison of the Two Models}
\label{subsec:results_comparison}

Our shifted-dipole result, $R=13.28^{+1.47}_{-1.07}\,\mathrm{km}$, has a central 68\% interval of $12.21$--$14.75\,\mathrm{km}$. For joint \emph{NICER}--\emph{XMM-Newton} analyses, \citet{Miller2021} reported $13.7^{+2.6}_{-1.5}\,\mathrm{km}$ and \citet{Riley2021} reported $12.39^{+1.30}_{-0.98}\,\mathrm{km}$. With the longer \emph{NICER} data set also used here, \citet{Dittmann24} obtained $12.92^{+2.09}_{-1.13}\,\mathrm{km}$ and \citet{Salmi_2024_j0740} obtained $12.49^{+1.28}_{-0.88}\,\mathrm{km}$. These intervals all overlap ours substantially. Thus, a temperature map generated from a physical magnetic and heating prescription can recover a radius inference broadly consistent with those obtained using flexible phenomenological hotspots.

The mass agreement has a different interpretation. Our $M=2.059^{+0.068}_{-0.069}\,M_\odot$ remains close to the informative radio prior, as do the masses inferred in previous PPM analyses. These analyses use the same or closely related Shapiro-delay information: \citet{Riley2021,Salmi_2024_j0740} use timing-informed mass--inclination constraints, whereas \citet{Miller2021,Dittmann24} broaden the mass uncertainty to $0.09\,M_\odot$ to allow for systematic error. Their agreement in mass is therefore largely expected and is not an independent validation of the hotspot model. The radius comparison instead tests the consequences of the different surface prescriptions conditional on that mass information.

Prior support also matters for the radius comparison. Our prior is uniform over $6$--$20\,\mathrm{km}$, whereas \citet{Riley2021,Salmi_2024_j0740} exclude $R>16\,\mathrm{km}$. \citet{Miller2021,Dittmann24} instead sample inverse compactness uniformly over $3.2\leq Rc^2/(GM)\leq8.0$, corresponding to approximately $9.83$--$24.57\,\mathrm{km}$ at $M=2.08\,M_\odot$, rather than using a mass-independent radius interval. Restricting the \citet{Dittmann24} samples to $R<16\,\mathrm{km}$ gives $R=12.76^{+1.49}_{-1.02}\,\mathrm{km}$. This illustrates how the upper prior support affects a high-radius tail, but it does not isolate the cause of differences with our result. Viewing priors, distance treatment, absorption, calibration, background treatment, and sampling also differ. In particular, we marginalize over the radio-timing distance distribution while fixing the absorption and instrumental normalizations. The posterior overlap thus establishes compatibility under the respective assumptions.

\begin{figure*}
\centering
\includegraphics[width=0.98\textwidth]{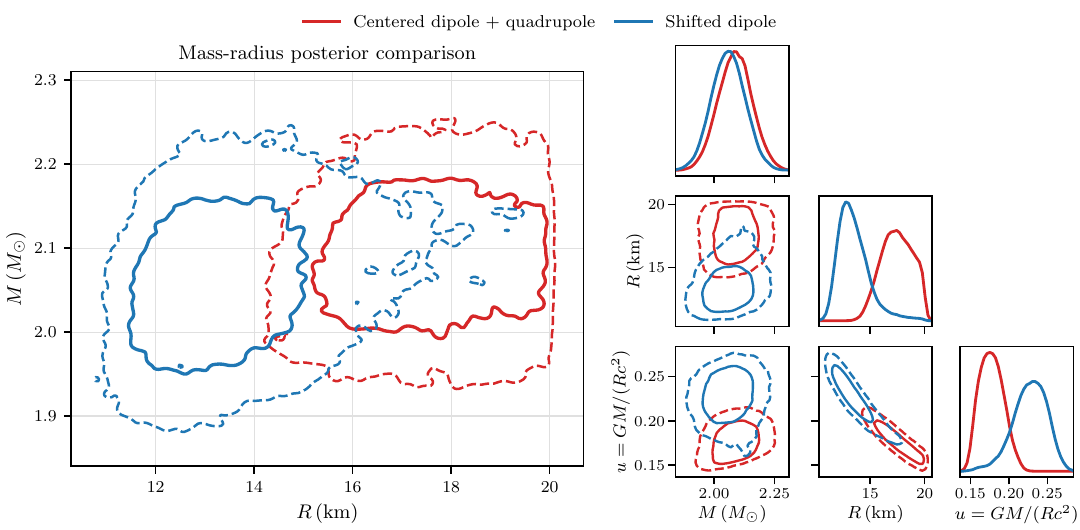}
\caption{Comparison of the stellar posteriors for the two primary magnetic models. Left: 68\% (solid) and 95\% (dashed) highest-posterior-density contours in the mass--radius plane. Right: marginalized mass, radius, and compactness distributions. Enforcing a centered source moves the inference toward a larger radius and lower compactness.\label{fig:mass_radius_comparison}}
\end{figure*}

The mass--radius comparison in Figure~\ref{fig:mass_radius_comparison} shows the central contrast between the two primary models. The shifted dipole is centered at $R\simeq13$--$14\,\mathrm{km}$ and $\compactness\simeq0.22$--0.23, whereas the centered dipole--quadrupole occupies $R\simeq16$--$19\,\mathrm{km}$ and $\compactness\simeq0.18$. Since the same mass prior, atmosphere, instrument response, background treatment, and heating prescription were used in both calculations, this difference arises from the magnetic geometry and its associated surface-temperature pattern. A model can therefore reproduce the phase and energy projections while shifting the inferred stellar compactness to compensate for a constrained surface geometry. This is precisely the degeneracy that physics-informed hotspot modeling is intended to reveal.

The comparison also suggests that the mass--radius inference is relatively insensitive to the detailed hotspot prescription once the surface model has sufficient geometric freedom to reproduce the observed emission pattern. The shifted-dipole result remains broadly consistent with previous phenomenological hotspot analyses despite imposing a substantially more restrictive connection between the magnetic field configuration. Within this class of sufficiently flexible models, differences in the inferred mass--radius distribution may therefore depend more strongly on choices such as the radius and viewing-angle priors, distance treatment, absorption, calibration, and background modeling than on the detailed hotspot parameterization itself. This stability does not extend to an overly restrictive physical model, as the centered dipole--quadrupole configuration cannot reproduce the required surface geometry without shifting the inference toward a substantially larger radius and lower compactness.

\begin{figure*}
\centering
\includegraphics[width=0.82\textwidth]{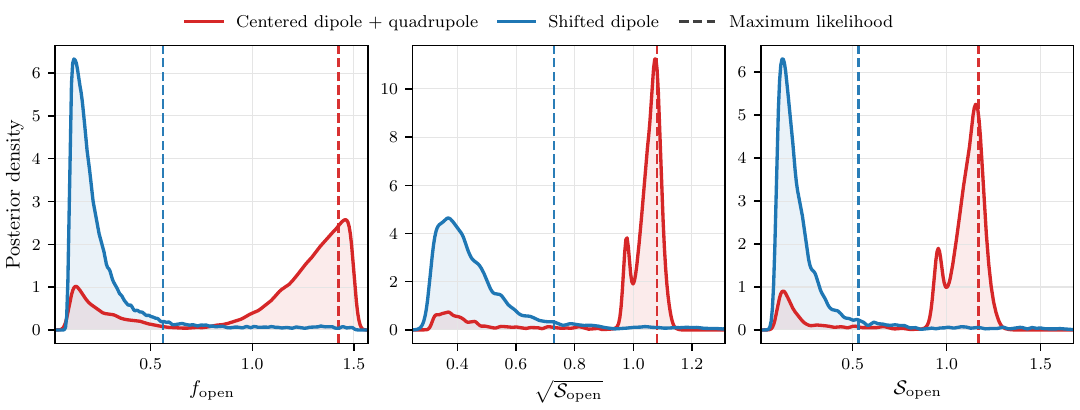}
\caption{Posterior comparison of $\fopen$ (left), the equivalent-area radius ratio $\sqrt{\Sopen}$ (center), and the open-footprint area ratio $\Sopen$ (right).  The latter two quantities are obtained by integrating the generally noncircular $\alpha<\alpha_0$ region and normalizing it to the area obtained for the same magnetic geometry with $\fopen=1$; see Equation~(\ref{eq:open_area_ratio}).  Solid curves are marginalized posterior densities from 4096 equal-weight samples, and dashed vertical lines mark the maximum-likelihood points whose complete vectors are listed in Table~\ref{tab:maxlike_parameters}.  The equivalent radius summarizes area only and does not encode the footprint shape or the return-current distribution within it.\label{fig:fopen_comparison}}
\end{figure*}

Figure~\ref{fig:fopen_comparison} makes the footprint contraction in the different models explicit. The sampled $\fopen$ is the multiplier of the flux cutoff and remains the natural quantity to compare with the force-free expectation. It is not directly a surface-area ratio in the presence of higher multipoles. The quantity $\Sopen$ measures the overall open-footprint area ratio. The radiating subset also depends on the quadrupole component, the return-current region, and the inferred heating normalizations. The posterior area ratios reported below are calculated directly from Equation~(\ref{eq:open_area_ratio}), without imposing the heating-activity mask or a cut on the resulting temperature.

\begin{figure*}
\centering
\includegraphics[width=0.98\textwidth]{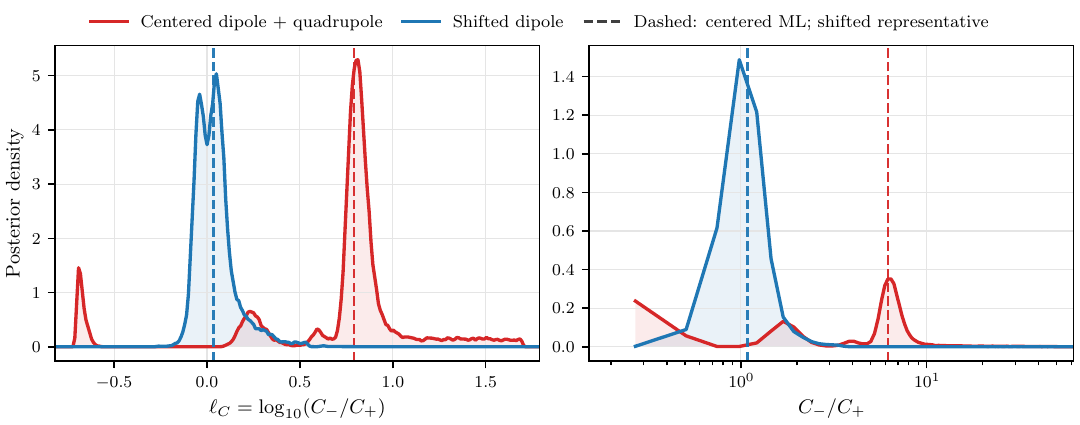}
\caption{Posterior comparison of the logarithmic (left) and linear (right) negative-to-positive-$\cos\theta'$ branch temperature-normalization ratio. Solid curves have the same meaning as in Figure~\ref{fig:fopen_comparison}. The dashed red line marks the centered-model global maximum, and the dashed blue line marks the timing-compatible shifted-dipole representative. Equal current-to-heat conversion corresponds to $\ell_C=0$ and $\Cminus/\Cplus=1$.\label{fig:cap_ratio_comparison}}
\end{figure*}

For the shifted dipole, $\Sopen=0.176^{+0.150}_{-0.059}$ and $\sqrt{\Sopen}=0.420^{+0.151}_{-0.077}$, close to but not identical to the small-angle estimates $\fopen$ and $\sqrt{\fopen}$. The centered dipole--quadrupole has $\fopen=1.275^{+0.165}_{-0.914}$ but only $\Sopen=1.111^{+0.068}_{-0.164}$. Its nonlinear quadrupolar contour makes the footprint area much less sensitive to the cutoff. Its median open region nevertheless covers 24.4\% of the stellar surface, compared with 1.1\% for the shifted dipole, explaining the broad map in Figure~\ref{fig:centered_temperature_map}. At the posterior median of $\fopen$, the shifted-dipole open footprint corresponds to $\approx25~\mathrm{km^2}$, whereas the nominal $\fopen=1$ footprint covers $\approx6.4\%$ of the surface, or $\approx140~\mathrm{km^2}$. For comparison, the two uniform hot regions of the phenomenological analyses, with angular radii $\zeta\approx0.10$--$0.15$~rad, together cover $1-\cos\zeta\approx0.5$--$1.1\%$ of the stellar surface, or $\approx10$--$20~\mathrm{km^2}$ at their inferred radii \citep{Miller2021,Riley2021,Dittmann24,Salmi_2024_j0740}. The physical and phenomenological models therefore require comparable emitting areas. Because the activity mask and the heating taper in Equation~(\ref{eq:heating}) remove part of the open footprint, the shifted-dipole value should be considered an upper bound on its X-ray-active area. Figures~\ref{fig:mass_radius_comparison}, \ref{fig:fopen_comparison}, and \ref{fig:cap_ratio_comparison} therefore make clear that displacement removes the large-radius and heating-asymmetry tensions of the centered model, but replaces them with a much smaller inferred effective open-field footprint.

The inference therefore separates a magnetic-flux question from a geometric-area question. Neither quantity directly measures a pair-production attenuation length or a kinetic gap height. About 92.0\% of the shifted-dipole posterior samples have $\Sopen<0.5$, corresponding to less than half the computed nominal polar-cap area. The timing-compatible representative has $\Sopen=0.173$ and lies within the contracted region favored by the posterior.
%In Section~\ref{subsec:discussion_cap} we make a restricted comparison with photon cascade to determine whether delayed pair conversion can plausibly account for the inferred contraction. 
In Section~\ref{subsec:discussion_cap}, we compare pair-cascade estimates as an illustrative check, while
Appendix~\ref{app:shifted_dq} tests whether additional magnetic structure can provide a solution toward the nominal force-free hotspot area.

Finally, Figure~\ref{fig:cap_ratio_comparison} compares the inferred negative-to-positive magnetic-branch heating asymmetry. The shifted dipole is concentrated around $\Cminus/\Cplus=1$, and the timing-compatible representative is also nearly symmetric, with $\Cminus/\Cplus=1.087$. The centered model is qualitatively different: its main solution requires the temperature normalization of one magnetic branch to be approximately six times that of the other. The shifted dipole therefore shows that the two observed pulses need not be produced by strongly different current-to-heat conversion between the two magnetic branches. Conversely, the centered field favors substantial branch-to-branch differences in the heating normalization.

The Bayesian evidence, the reproduction of the pulse-profile data, the nearly symmetric heating between the two magnetic branches, and the mass--radius posterior identify the shifted dipole as our preferred physical model for PSR~J0740+6620. It reproduces the observations without requiring the large radius or strong branch-to-branch heating asymmetry of the centered dipole--quadrupole model, while yielding a mass--radius posterior broadly consistent with previous phenomenological analyses. Both its marginalized posterior and timing-compatible representative solution require an effective open-field footprint smaller than the nominal force-free scale calibrated for a centered dipole. The shifted-dipole model therefore removes the main geometrical and heating tensions of the centered configuration while exposing a different unresolved piece of the physics: the mapping from magnetospheric current to observable X-ray emission. This may require a more complete surface-emission and heating prescription, potentially including spatially concentrated heating associated with the unresolved return-current sheet.

\section{Discussion}
\label{sec:discussion}

The two primary models considered in the main text differ substantially in their inferred mass--radius distributions and in the physical conditions required to reproduce the X-ray data. The centered dipole--quadrupole matches the displayed projections without obvious residual structure {but} favors a large radius, with its posterior extending to the upper prior boundary, together with a broad footprint and strongly asymmetric heating. The shifted dipole gives a radius consistent with previous analyses and nearly symmetric heating, making it the more natural model. Its remaining problem is specific: the inferred effective open-field footprint is much smaller than the nominal force-free polar cap. The timing-compatible representative gives $\fopen\approx0.194$, close to the marginalized posterior median $\fopen\approx0.19$. This contraction may indicate missing physics in our mapping from open-zone currents to surface emission. We next consider pair-cascade microphysics and surface heating.  In addition, we discuss constraints on the inferred viewing angle derived under the assumption of spin--orbit alignment.

\subsection{Physical Explanations for the Cap Contraction}
\label{subsec:discussion_cap}

The shifted-dipole posterior requires $\fopen=0.187^{+0.174}_{-0.062}$, corresponding to an open footprint of about $1\%$ of the stellar surface. As shown in Section~\ref{subsec:results_comparison}, this area is comparable to previous phenomenological analyses of the same data assuming uniform hot regions, so the small emitting area is a robust property of the observations rather than of our surface prescription. In our model, $\fopen$ is the effective flux cutoff relative to the nominal flat-spacetime value $\alpha_{0,\mathrm{FF}}$. We therefore discuss which physical effects can reduce the effective open footprint below its nominal size, considering in turn general relativity, delayed pair conversion, and alternative pair-production channels, before discussing the implications for the surface magnetic field.

\emph{General relativity.} As shown in Section~\ref{sec:template}, the Schwarzschild dipole concentrates the surface flux by the factor $\Delta_1(\compactness)$ of Equation~(\ref{eq:delta1}). For the shifted-dipole posterior compactness, $\compactness=0.229^{+0.020}_{-0.023}$, the relativistic nominal footprint therefore corresponds to $\fopen\simeq1/\Delta_1=0.65^{+0.04}_{-0.03}$. No relativistic solution exists for a shifted dipole, but a comparable near-surface flux concentration is expected.

\emph{Delayed pair conversion.} In the standard picture, particles accelerated in a gap above the polar cap emit curvature photons that convert to $e^\pm$ pairs in the magnetic field \citep{Erber1966,Timokhin2013}. Because these photons propagate approximately tangent to the field at emission while the field lines continue to curve, pairs produced after a finite propagation distance populate field lines whose footpoints lie inside the original emission footpoints, closer to the polar cap center. Returning particles could then heat a region smaller than the full polar cap. Consider a photon emitted along the field at the cap rim, at magnetic colatitude $\theta_{\rm pc}$ and cylindrical radius $\varpi\simeq R\theta_{\rm pc}$. The field-line tangent makes an angle $3\theta_{\rm pc}/2$ with the magnetic axis, so after a distance $s$ the photon lies at $\varpi_\gamma\simeq\theta_{\rm pc}(R+3s/2)$ and $r_\gamma\simeq R+s$. Mapping this point back to the surface along a dipole field line, for which $\sin^2\theta_{\rm B}/r$ is constant, gives the ratio of the footprint radius of the conversion field line to the original cap radius,
\begin{equation}
g(x)=\frac{1+3x/2}{(1+x)^{3/2}},\qquad x\equiv\frac{s}{R},\qquad \fopen=g^2(x).
\label{eq:ballistic_contraction}
\end{equation}
The conversion distance is limited by the dipolar decline of the field. Photons convert once the quantum parameter $\chi\simeq\epsilon(B/B_Q)\sin\psi$ reaches $\approx0.1$--$0.2$, where $\epsilon$ is the photon energy in units of $m_ec^2$ and $\psi\simeq s/\rho_c$ is the angle between the photon and the field after a distance $s$, with $\rho_c$ the field-line curvature radius \citep{Timokhin2013}. With $B\propto(1+x)^{-3}$, $\chi\propto x(1+x)^{-3}$ peaks at $x=1/2$ and declines beyond it, so a photon that has not converted by $s\approx R/2$ escapes. Delayed conversion can therefore contract the footprint by at most $g^2(1/2)\simeq0.91$. This bound is independent of the uncertain gap parameters, which determine only whether conversion occurs at all. We do not apply Equation~(\ref{eq:ballistic_contraction}) to the quadrupolar models, whose non-circular footprints violate its local dipolar mapping.

Whether rim curvature photons convert at all is nevertheless relevant for the pair supply of PSR~J0740+6620. The conventional vacuum spin-down estimate for $\nu=346.5319965~\mathrm{Hz}$ and $\dot{\nu}=-1.46387\times10^{-15}~\mathrm{s^{-2}}$ \citep{Fonseca_21}, $B_{\rm sd}=3.2\times10^{19}(P\dot P)^{1/2}~\mathrm{G}=1.9\times10^8~\mathrm{G}$ for a $10~\mathrm{km}$ star with $I=10^{45}~\mathrm{g\,cm^2}$, corresponds to a dipole moment $\mu\simeq1.9\times10^{26}~\mathrm{G\,cm^3}$ and hence, for $R=13.3~\mathrm{km}$, a polar surface field $B_{\rm p}=2\mu/R^3\simeq1.6\times10^8~\mathrm{G}$. Curvature losses limit the primaries to a Lorentz factor $\gamma_{\rm rad}$ well below that of the full polar-cap potential, $\gamma_{\rm pc}\simeq5.9\times10^8$ \citep{Ruderman1975}, obtained by balancing acceleration against radiation reaction \citep[e.g.,][]{YeChen2025},
\begin{equation}
eE_\parallel=\frac{2e^2}{3\rho_c^2}\gamma_{\rm rad}^4 .
\label{eq:gamma_rad}
\end{equation}
For the vacuum-gap field $E_\parallel\simeq4\pi\rho_{\rm GJ}r_{\rm pc}=2\Omega B_{\rm p}r_{\rm pc}/c$, with $r_{\rm pc}\simeq R(R/\RLC)^{1/2}\simeq4.1~\mathrm{km}$ and $\rho_c\simeq(4/3)(R\RLC)^{1/2}\simeq5.7\times10^6~\mathrm{cm}$ near the cap rim, this gives $\gamma_{\rm rad}\simeq3.2\times10^7$ and a characteristic curvature-photon energy $\epsilon_{\rm curv}=3\hbar\gamma_{\rm rad}^3/(2\rho_c m_ec)\simeq3.2\times10^5$. Reaching $\chi\simeq0.2$ at the peak $s=R/2$ instead requires $\epsilon\gtrsim1.6\times10^6$, a factor of $\approx5$ higher. Because a screened gap would give lower $E_\parallel$, this vacuum-gap estimate is optimistic: rim curvature photons at the characteristic energy escape without converting. We verified this by tracing rim photons through the timing-compatible representative  shifted-dipole field and integrating the magnetic pair-production attenuation coefficient of \citet{Erber1966} along straight paths. Although the rim field varies by a factor of $\approx4$ around each displaced cap, photons at the characteristic energy escape, and those energetic enough to convert yield $\fopen\geq0.91$.

\emph{Other photon sources and conversion processes.} Pair production involves two steps: the production of a $\gamma$-ray, by curvature radiation or by inverse-Compton scattering (ICS) of surface X-rays, and its conversion, either in the magnetic field (one-photon conversion) or through collision with a soft X-ray photon (two-photon conversion) \citep{Harding_2002,Salmi2026}. Two-photon conversion depends steeply on the surface temperature. For the $kT\approx0.09~\mathrm{keV}$ hotspots of PSR~J0740+6620, integrating the Breit--Wheeler cross section over the radiation field of a hot region of radius $1.5$--$4~\mathrm{km}$ gives an optical depth $\tau\lesssim0.1$ for outward-moving $\gamma$-rays of any energy, compared with $\tau\approx1$ at $kT\approx0.3~\mathrm{keV}$. This is consistent with the finding of \citet{Salmi2026} that one-photon conversion can dominate below $kT\approx0.1~\mathrm{keV}$. Where two-photon conversion does occur, it is confined to within about one hotspot radius of the surface, where the soft-photon density and collision angles are largest, and therefore produces no significant contraction. ICS instead provides an alternative photon source. In the Klein--Nishina regime, the scattered photons carry a large fraction of the particle energy, well above $\epsilon_{\rm curv}$, and convert magnetically within $s\ll R$, again close to their emission footpoints. The ICS rate, however, is also proportional to the soft-photon density and is correspondingly lower at this surface temperature than in hotter millisecond pulsars. Neither process therefore contracts the footprint.

\emph{Implications for the surface field.} Together, the relativistic normalization and delayed conversion can contract the effective footprint only to $\fopen\approx0.6$, whereas the shifted-dipole posterior favors $\fopen\approx0.19$. Two independent considerations suggest that part of this discrepancy may reflect magnetic structure near the surface that is stronger than a displaced dipole. First, if the open flux crosses the surface where the field is enhanced by a factor $b\equiv B_{\rm surf}/B_{\rm dip}$, flux conservation reduces the footprint by $1/b$. The posterior area ratio $\Sopen=0.176^{+0.150}_{-0.059}$ then requires $b\approx6$ ($3$--$9$) for a flat-spacetime dipole, or $b\approx4$ ($2$--$6$) after the relativistic normalization. Second, pairs are required both for the coherent radio emission of PSR~J0740+6620 and for the return-current heating of its polar caps that produces the thermal X-rays modeled here. At the dipolar field strength, however, the estimate above places the star near the death line for curvature-photon pair production, while two-photon conversion and ICS are inefficient at its surface temperature. Unless the pairs are supplied elsewhere in the magnetosphere, the presence of both radio pulsations and heated polar caps therefore also points to a stronger surface field. Because $\epsilon_{\rm curv}\propto B^{3/4}$ for a radiation-reaction-limited vacuum gap while the conversion threshold scales as $B^{-1}$, an enhancement $b\gtrsim2.5$ would allow curvature photons to convert near the surface, consistent with the enhancement required by the footprint. Small-scale surface fields have long been invoked for millisecond pulsars near or below the curvature death line \citep{Harding_2002,Harding2022}, and multipolar or double-dipole configurations have been proposed for PSR~J0740+6620 itself \citep{Petri_2025_pulsar,Taiyebah2026}. The shifted dipole--quadrupole extension in Appendix~\ref{app:shifted_dq} provides a concrete example: its maximum-likelihood solution reaches $\fopen\simeq1$ because the quadrupole itself concentrates the surface flux, although the present data do not require this additional freedom.

In summary, none of the mechanisms considered here, namely the general-relativistic correction of the open flux, the delayed magnetic conversion of curvature photons, and two-photon conversion or inverse-Compton photon production, contracts the effective footprint below $\fopen\approx0.6$, a factor of $\approx3$ above the posterior median. A stronger near-surface field is one plausible explanation of the remaining contraction, but not the only one. Because $\fopen$ constrains only the combined effect of the open-flux normalization, the near-surface magnetic geometry, and the heating efficiency across the cap, the discrepancy may equally reflect our simplified heating prescription, which is by no means complete.

\subsection{Heating from the Return-Current Sheet}

\label{subsec:discussion_current_sheet}

An obvious omission of our polar cap heating model is the return current sheet concentrated at the separatrix between the open and closed zones. The surface-heating prescription used here includes only the resolved current components of the smooth force-free polar-cap solution. In the force-free approximation adopted by our model, the separatrix current sheet is an infinitely thin layer of discontinuity where the current density is effectively infinite. In reality, the width is set by the kinetic scales of the plasma supporting the current, likely on the order of a few plasma skin depths. Neither the width nor the charge composition of the separatrix current sheet is well understood, and must be established by future magnetospheric calculations with sufficient resolution near the star. The pair-production mechanism that supplies the sheet, and hence the fraction and energy of particles returning to the surface, are likewise uncertain. It is therefore plausible that current-sheet precipitation could contribute heating comparable to, or even greater than, that from the resolved direct-current components included here, changing the emitting area and temperature structure near the cap boundary. Quantifying this contribution will require simulations that resolve the current sheet, its pair supply, and the deposition of returning-particle energy at the surface, which is far beyond the scope of this paper. Including current-sheet heating could modify the hotspot temperature distribution near the cap boundary, potentially producing a more edge-brightened structure than in the present smooth-current prescription. The thinness of the structure may contribute to the origin of the significantly smaller $f_\mathrm{open}$ that we infer from the data. Determining the contribution of return-current heating to the inferred X-ray morphology is therefore a crucial next step in this line of research. 

\subsection{Polar-cap Heating Asymmetry and Quadrupole Influence}
\label{subsec:discussion_heating}

Because the magnetic model predicts the relative return-current pattern in both branches, their fitted normalization ratio tests the current-to-heat conversion after accounting for differences in the incident current. A temperature contrast between phenomenological spots alone could not make this distinction, because it would conflate the current supplied to each spot with the efficiency of converting that current into heat. For the shifted dipole, the inferred ratio is $\Cminus/\Cplus=1.074^{+0.233}_{-0.183}$, fully consistent with a common current-to-temperature normalization in the negative- and positive-$\cos\theta'$ branches. However, the centered dipole--quadrupole fit behaves differently. It favors $\Cminus/\Cplus=6.36^{+1.65}_{-4.19}$, and only 7.9\% of its posterior allows the two normalizations to agree within a factor of two. Thus, once the magnetic source is allowed to be displaced, the observed pulse asymmetry does not require intrinsically different current-to-heat conversion efficiencies in the two branches. Keeping the source centered instead requires a strong branch-to-branch heating asymmetry in the statistically disfavored centered dipole--quadrupole model.

The quadrupole modifies the heating pattern through both the surface flux and the field-aligned current: $q_\alpha$ reshapes the open footprints, whereas $q_\Lambda$ redistributes the return current within them. Among the configurations tested, the shifted dipole provides a better-supported description of the observations than the centered dipole--quadrupole model. As shown in Appendix~\ref{app:shifted_dq}, adding a quadrupole to the shifted geometry remains a viable way to modify the footprint, and a high-likelihood branch approaches the nominal force-free cap scale. However, the broader posterior and lower evidence show that the present observations do not require this additional magnetic freedom.
\subsection{Viewing Inclination and Spin--Orbit Alignment}
\label{subsec:discussion_inclination}

For the shifted dipole, the posterior samples give $\obsinc=70.20^{+13.61}_{-18.45}$ degrees, with a central 95\% interval of $31.31^\circ$--$89.05^\circ$. The posterior therefore extends to the nearly edge-on geometry inferred from radio timing, and the Shapiro-delay inclination lies within its 95\% interval. The median is $17.36^\circ$ below the radio value. We use the global maximum to report $\ln\mathcal L_{\max}$, but its parameter vector is not representative of the marginalized solution. It has $\obsinc=17.86^\circ$, whereas only 2.1\% of the shifted-dipole posterior has $\obsinc<30^\circ$. Its radius and open-flux cutoff also lie in the upper posterior tails. We therefore do not use it for the physical illustration of the preferred model.

For a broader comparison with radio-timing-informed PPM, the interval $82.5^\circ$--$90^\circ$ is the folded version of the $\pm5^\circ$ viewing window used by \citet{Miller2021}. The shifted-dipole posterior has substantial support near the radio-timing inclination: about 20\% of the posterior lies within $82.5^\circ$--$90^\circ$, while the radio value itself lies within the 95\% credible interval. The centered-model median, $\obsinc\simeq39.9^\circ$, would imply at least $48^\circ$ of spin--orbit misalignment if interpreted literally; together with its large radius and unequal heating, we interpret this more naturally as model-dependent compensation than as evidence for a large spin tilt.

For the representative shifted-dipole solution, we selected the highest-likelihood stored sample in the radio 68\% orbital-inclination interval, $87.38^\circ\leq\orbinc\leq87.73^\circ$, which we apply to $\obsinc$ under the assumption of spin--orbit alignment. We reevaluated its joint \emph{NICER}--\emph{XMM-Newton} likelihood with the production forward model. This sample has $\obsinc=87.5044^\circ$, $M=2.0253\,M_\odot$, $R=13.2630\,\mathrm{km}$, and $\ln\mathcal L=-21863.08377$, compared with $-21862.10986$ at the global maximum, corresponding to a likelihood ratio of 0.378. On the $256\times256$ map, the local maximum and X-ray-active-area mean temperatures are $T_{\max}=1.414\times10^6~\mathrm{K}$ and $\langle T\rangle_{\rm X}=7.758\times10^5~\mathrm{K}$, using the definition in Table~\ref{tab:maxlike_parameters}.

The representative has $\maginc=87.29^\circ$, a substantial displacement $|\bm s|/R\simeq0.567$, nearly equal branch normalization $\Cminus/\Cplus=1.087$, $\fopen=0.1942$, and $\Sopen=0.1726$. At this small likelihood cost, it retains the mass--radius in a reasonable range, nearly symmetric heating, and contracted footprint of the marginalized shifted-dipole solution. Here the ``timing-compatible'' assumes spin--orbit alignment automatically. Shapiro-delay timing constrains $\orbinc$, not $\obsinc$ directly. We did not apply the same display selection to the other magnetic families. We retain the centered-model global maximum to show its best fit. A timing-informed model comparison would require the same radio-informed prior or likelihood for every model and new posterior and evidence calculations.

If the X-ray hotspots are shaped by displaced or multipolar fields near the surface, the inferred magnetic inclination describes this near-surface magnetic geometry and need not equal the large-scale dipole obliquity, as also suggested by the multiwavelength comparison of \citet{Petri_2025_pulsar}, testing this interpretation requires a self-consistent global multiwavelength model.

\section{Conclusions}
\label{sec:conclusion}

In this work, we have jointly inferred the mass, radius, and magnetic geometry of PSR~J0740+6620 using physically motivated force-free-current temperature maps and GPU-accelerated PPM of \emph{NICER} and \emph{XMM-Newton} data. The shifted dipole is our preferred physical model and gives $M=2.059^{+0.068}_{-0.069}\,M_\odot$ and $R=13.28^{+1.47}_{-1.07}\,\mathrm{km}$. Its radius posterior substantially overlaps those from previous phenomenological analyses \citep{Miller2021,Riley2021,Dittmann24,Salmi_2024_j0740}, showing that the inferred mass--radius region persists when freely parameterized hotspots are replaced by a temperature map tied directly to the magnetic field and magnetospheric current. 

The two primary magnetic prescriptions differ in both statistical support and physical implications. The centered dipole--quadrupole favors $R=17.46^{+1.57}_{-1.48}\,\mathrm{km}$, with the radius posterior extending toward the upper prior boundary, together with a broad footprint and a branch-heating normalization ratio near six. The shifted dipole is favored by $\Delta\ln Z=5.654\pm0.177$ and reproduces the data with nearly equal heating normalization between the two magnetic branches and without requiring the large-radius solution of the centered model. The comparison shows that the mass--radius inference can remain relatively stable once the surface model has sufficient geometric freedom, whereas a more restrictive magnetic prescription can shift the inferred stellar compactness to compensate for the allowed surface geometry. These conclusions remain conditional on the adopted priors and nuisance-parameter treatment, including the fixed absorption and nominal instrumental normalizations.

The preferred shifted-dipole model nevertheless exposes a specific unresolved tension. At the global maximum, the effective footprint is already contracted relative to the nominal force-free scale, with $\fopen=0.561$ and $\Sopen=0.531$, while the marginalized posterior favors an even stronger contraction, $\fopen=0.187^{+0.174}_{-0.062}$ and $\Sopen=0.176^{+0.150}_{-0.059}$. The global maximum-likelihood viewing angle lies far from the nearly edge-on inclination inferred from Shapiro delay, but this point is not representative of the full posterior. The shifted-dipole posterior extends through the radio inclination, and a stored solution within the Shapiro-delay 68\% interval has a likelihood loss of only $\Delta\ln\mathcal L=0.974$, while retaining a similar mass, radius, and displaced magnetic geometry. This radio-compatible solution requires $\fopen=0.194$ and $\Sopen=0.173$. Agreement with the independent viewing geometry therefore does not remove the small-footprint tension and, for this solution, amplifies it.

The exploratory shifted dipole--quadrupole model presented in the appendix admits a high-likelihood branch with a footprint near the nominal force-free scale, but its lower evidence and broad posterior suggests that we do not require this additional magnetic structure to explain the data. Our simple delayed pair-conversion estimate also produces some contraction but is insufficient to explain its inferred magnitude. The remaining discrepancy therefore points to missing physics in the mapping from magnetospheric current to observable X-ray emission rather than requiring the shifted geometry itself to be rejected. More complete kinetic calculations of pair production, particle return, and surface energy deposition, including the unresolved separatrix return-current sheet, are needed to determine this mapping. An alternative route to explain the apparently too small heated region may be forcing a magnetic component close to the surface, similar to~\citet{Petri_2025_pulsar}, but a fully self-consistent surface heating prescription is still a required step. Radio and gamma-ray constraints, together with more complete treatments of surface emission and nuisance parameters, can further test whether the inferred magnetic geometry and mass--radius result persist.

The same framework also provides a path toward genuinely multiwavelength inference of pulsar magnetospheres, in the spirit of e.g.\ \citet{Chen_2020}, \citet{Kalapotharakos_2021}, \citet{Olmschenk:2025eie}, and \citet{Taiyebah2026}. Because the X-ray surface temperature map is tied to an explicit magnetic geometry and current distribution, the magnetic-field geometry can be fitted simultaneously to radio polarization and gamma-ray light curves, rather than treating each observable with an independent geometric model. Applying this approach to a larger sample of millisecond pulsars will show whether displaced or multipolar fields are common, whether the small-footprint tension found here is generic, and how strongly neutron-star mass--radius inference depends on the choice of magnetospheric model. With improved current and surface-heating prescriptions, future multiwavelength pulse-profile modeling can therefore become a tool for jointly measuring neutron-star interior structure through the mass and radius and probing the magnetosphere that produces the observed emission across different wavebands.

\section{acknowledgments}
The authors thank Tianzhe Zhou for valuable discussions and comments that substantially improved this work. A.C.\ and C.H.\ acknowledge support from NSF grant AST-2308111. A.C.\ additionally acknowledges support from NASA grant 80NSSC24K1095.
T.S.\ acknowledges funding by the Research Council of Finland grant 368807 and the Centre of Excellence in Neutron-Star Physics (grant 374063).

The authors used generative artificial-intelligence tools, including OpenAI Codex and Anthropic Claude, to assist with numerical cross-checks, code development, and language editing during the analysis and preparation of this manuscript. All scientific results, interpretations, and final manuscript content were independently reviewed and verified by the authors.

\appendix
\section{Shifted Dipole plus Quadrupole Extension}
\label{app:shifted_dq}

\begin{deluxetable*}{lrrrr}
\tablecaption{Comparison with the Shifted Dipole--Quadrupole Extension\label{tab:appendix_evidence}}
\tablehead{
\colhead{Magnetic model} & \colhead{$k$} & \colhead{$\ln\mathcal{L}_{\max}$} &
\colhead{$\ln Z$} & \colhead{$\Delta\ln Z$}}
\startdata
Shifted dipole & 12 & $-21862.110$ & $-21890.738\pm0.118$ & $0$ \\
Shifted dipole--quadrupole & 14 & $-21860.778$ & $-21892.541\pm0.080$ & $-1.803\pm0.142$ \\
\enddata
\tablecomments{Evidence differences are relative to the shifted dipole. The extension attains the largest maximum likelihood among the calculated models, but its extra prior volume is not rewarded by the evidence.}
\end{deluxetable*}

\begin{deluxetable*}{lc}
\tabletypesize{\scriptsize}
\tablecaption{Maximum-Likelihood Configuration of the Exploratory Extension}\label{tab:maxlike_extension}
\tablehead{\colhead{Parameter or quantity} & \colhead{Shifted dipole--quadrupole}}
\startdata
\cutinhead{Sampled parameters}
$x_s/R$ & $-0.03273$ \\
$y_s/R$ & $0.1380$ \\
$z_s/R$ & $-0.6256$ \\
$q_\alpha$ & $0.1409$ \\
$q_\Lambda$ & $-0.4315$ \\
$M\ (M_\odot)$ & $2.072$ \\
$R\ (\mathrm{km})$ & $14.13$ \\
$\ell_{\rm floor}$ & $4.403$ \\
$\cos\maginc$ & $0.1782$ \\
$\cos\obsinc$ & $0.9282$ \\
$\ell_T$ & $0.3510$ \\
$\ell_C$ & $0.007124$ \\
$\fopen$ & $1.002$ \\
$\phi_0$ (cycles) & $0.7029$ \\
\cutinhead{Derived quantities}
$\compactness$ & $0.2165$ \\
$\maginc$ (deg) & $79.74$ \\
$\obsinc$ (deg) & $21.85$ \\
$\Cminus/\Cplus$ & $1.017$ \\
$\Sopen$ & $1.003$ \\
$\sqrt{\Sopen}$ & $1.001$ \\
$T_{\max}\ (10^6\,\mathrm{K})$ & $2.396$ \\
$\langle T\rangle_{\rm X}\ (10^6\,\mathrm{K})$ & $0.8585$ \\
\enddata
\tablecomments{Sampled entries are the maximum-likelihood parameter values for the exploratory extension shifted dipole--quadrupole model, all values are rounded here to four significant figures.  A blank entry denotes a parameter absent from that magnetic family.  Fixed inputs such as $D$, $N_{\rm H}$, responses, and stellar spin are not repeated.  Detector backgrounds are not components of these sampled vectors; the two-dimensional count figures add the channel-wise background that maximizes the likelihood conditional on the listed sample.}
\end{deluxetable*}

As a secondary test, we allowed the magnetic-source displacement and quadrupolar structure to vary simultaneously. The shifted dipole--quadrupole model shifts the dipole and quadrupole together and allows all five magnetic quantities $(x_s/R,y_s/R,z_s/R,q_\alpha,q_\Lambda)$ to vary, giving 14 sampled parameters. This extension asks whether moderate multipolar structure can preserve the successful shifted model geometry while resolving the small-footprint tension of the shifted dipole.

Table~\ref{tab:appendix_evidence} compares the extension with the shifted dipole. The shifted dipole--quadrupole reaches the largest maximum likelihood of the calculated models, improving $\ln\mathcal L_{\max}$ by 1.33 relative to the shifted dipole. The improvement is small compared with the two additional degrees of freedom, and its Bayesian evidence is lower than that of the shifted dipole by $1.803\pm0.142$. The present data therefore do not require the quadrupole, even though a particular high-likelihood solution is physically informative.

\begin{figure*}
\centering
\includegraphics[width=0.98\textwidth]{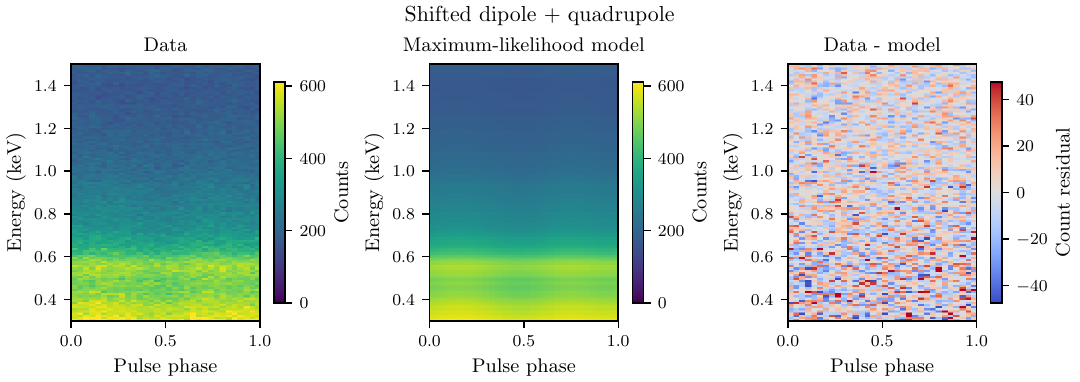}
\caption{As in Figure~\ref{fig:shifted_counts}, but for the shifted dipole--quadrupole extension. The parameter vector is listed in Table~\ref{tab:maxlike_extension}.\label{fig:shifted_dq_counts}}
\end{figure*}

\begin{figure*}
\centering
\includegraphics[width=0.92\textwidth]{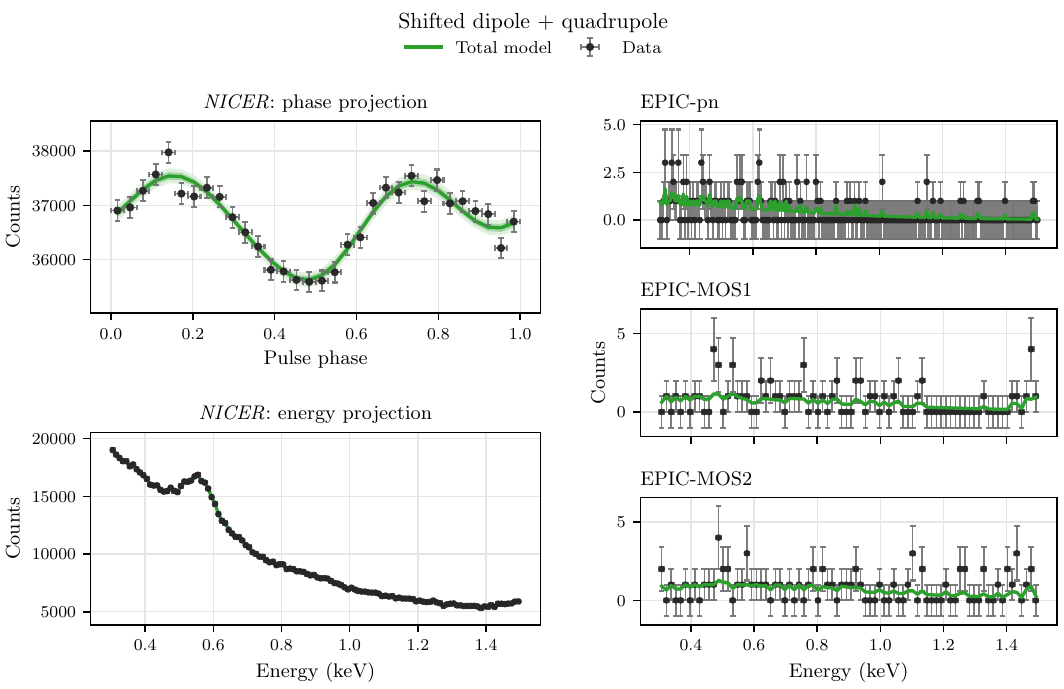}
\caption{As in Figure~\ref{fig:centered_instrument_projections}, but for the shifted dipole--quadrupole extension.\label{fig:shifted_dq_instrument_projections}}
\end{figure*}

Figures~\ref{fig:shifted_dq_counts} and \ref{fig:shifted_dq_instrument_projections} show that the phase--energy residuals and the \emph{NICER} and EPIC projections are as satisfactory as those of the two primary models.

The mass--radius posterior returns to the shifted-dipole solution, with $M=2.062^{+0.069}_{-0.067}\,M_\odot$, $R=13.54^{+2.50}_{-1.23}\,\mathrm{km}$, and median $\compactness=0.224$. The radius posterior has a main mode near $13\,\mathrm{km}$ and a noticeable high-radius shoulder; only 3.8\% of its probability lies above $18\,\mathrm{km}$. This shoulder widens the upper uncertainty but does not reproduce the dominant $17$--$19\,\mathrm{km}$ solution of the centered model. The median angles, $\maginc\simeq80^\circ$ and $\obsinc\simeq68^\circ$, are also close to the shifted-dipole geometry.

As shown in Figure \ref{fig:shifted_dq_corner}, the two additional magnetic parameters are not significant in their one-dimensional marginal distributions. The figure also shows their covariances with $\fopen$ and the cap-normalization parameters. The posterior values $q_\alpha=-0.14^{+0.56}_{-3.18}$ and $q_\Lambda=-0.070^{+0.377}_{-0.293}$ both include zero. The maximum-likelihood solution instead has $q_\alpha=0.141$ and $q_\Lambda=-0.431$. It should therefore be described as a particular high-likelihood quadrupolar distortion, not as a posterior measurement of a nonzero quadrupole.

\begin{figure*}
\centering
\includegraphics[width=0.78\textwidth]{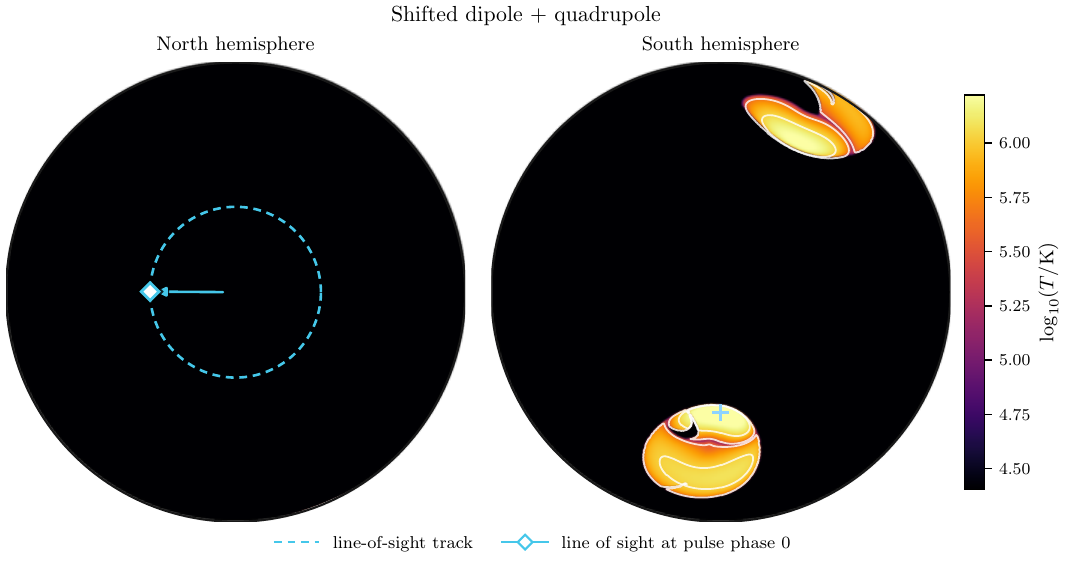}
\caption{Maximum-likelihood shifted dipole--quadrupole temperature map, shown as in Figure~\ref{fig:shifted_temperature_map}. The map preserves the compact two-region organization of the shifted dipole while adding quadrupole-driven current structure. The parameter vector is listed in Table~\ref{tab:maxlike_extension}.\label{fig:shifted_dq_temperature_map}}
\end{figure*}

\begin{figure*}
\centering
\includegraphics[width=0.98\textwidth]{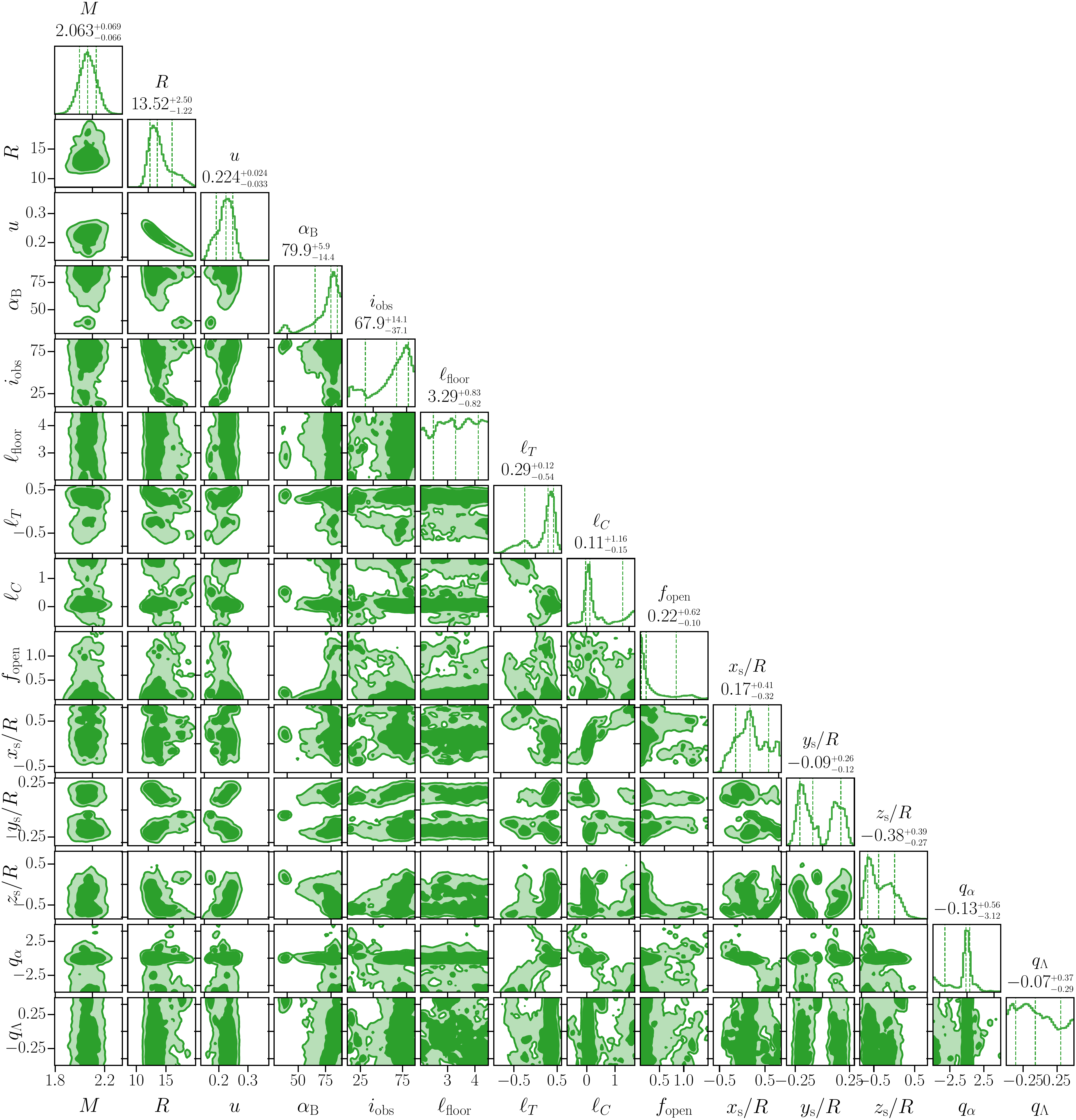}
\caption{Selected parameter posterior distributions for the shifted dipole--quadrupole extension. Contours and titles follow Figure~\ref{fig:shifted_corner}. The broad and multimodal quadrupole, $\fopen$, and cap-normalization marginals distinguish the posterior from the nominal-footprint maximum-likelihood point.\label{fig:shifted_dq_corner}}
\end{figure*}

After aligning the arbitrary longitude of the shifted-dipole and shifted dipole--quadrupole maps, the maximum-likelihood temperature distribution in Figure~\ref{fig:shifted_dq_temperature_map} retains the compact two-region organization of the shifted dipole. The quadrupolar current adds internal and boundary structure that would be difficult to represent with a simple phenomenological hotspot.

The most important feature of the extension is its ability to resolve the small-footprint tension at a particular high-likelihood point. At the maximum-likelihood solution, $\fopen=1.002$, $\Sopen=1.003$, and $\Cminus/\Cplus=1.017$. The nominal force-free polar-cap size and nearly equal hemispheric heating can therefore reproduce the data. This contrasts with the maximum-likelihood shifted dipole, for which $\Sopen=0.531$, and demonstrates that higher multipolar structure can provide a magnetic route to removing the cap-size tension.

The full posterior is less decisive than this best-fitting configuration. Its marginalized distribution gives $\fopen=0.222^{+0.625}_{-0.097}$, and 12.6\% of the samples lie in $0.8<\fopen<1.2$, compared with 2.6\% for the shifted dipole. After the nonlinear surface mapping, $\Sopen=0.279^{+0.683}_{-0.152}$ and 28.5\% of the area posterior lies within 20\% of the nominal footprint. The cap-ratio posterior has a near-symmetric mode and a long high-contrast tail; its median is 1.31, its maximum-likelihood value is 1.017, and 61.8\% of the posterior lies within a factor of two of equal heating. The data therefore admit, but do not isolate, the physically attractive nominal-area and symmetric-heating branch.

The extension also contains a contracted branch: a small-cutoff point with $\fopen=0.298$ lies only $\Delta\ln\mathcal L=1.51$ below the maximum. The extension therefore contains two physically distinct ways to explain the data: a nominal force-free-area branch with appreciable multipolar structure and a contracted branch with a smaller quadrupolar contribution. Their likelihoods are close enough that their relative posterior weights are controlled mainly by the available parameter volume.

\begin{figure*}
\centering
\includegraphics[width=0.98\textwidth]{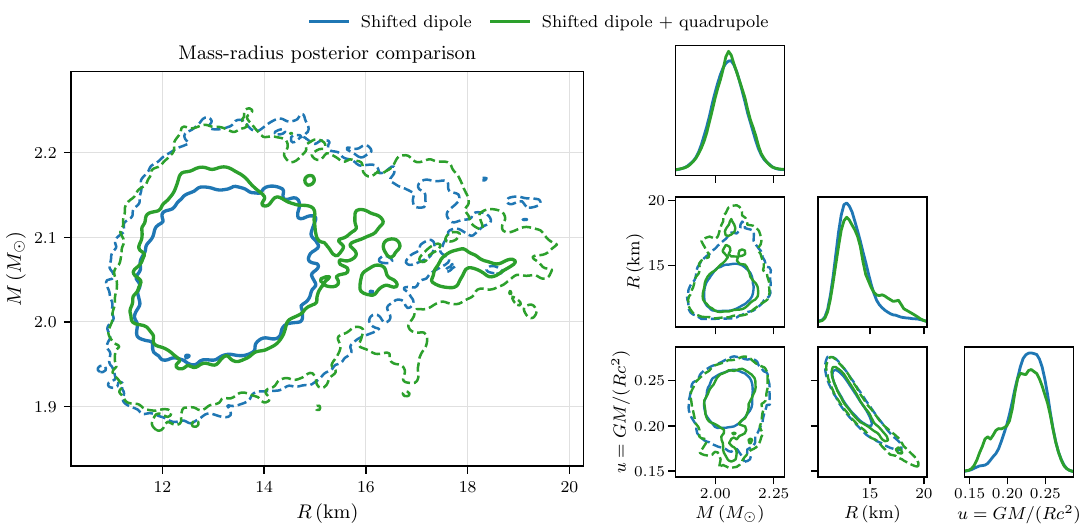}
\caption{Mass--radius and compactness comparison of the shifted dipole with its quadrupolar extension. The extension preserves the principal shifted-dipole stellar solution but develops a broader high-radius shoulder.\label{fig:appendix_mass_radius_comparison}}
\end{figure*}

\begin{figure*}
\centering
\includegraphics[width=0.98\textwidth]{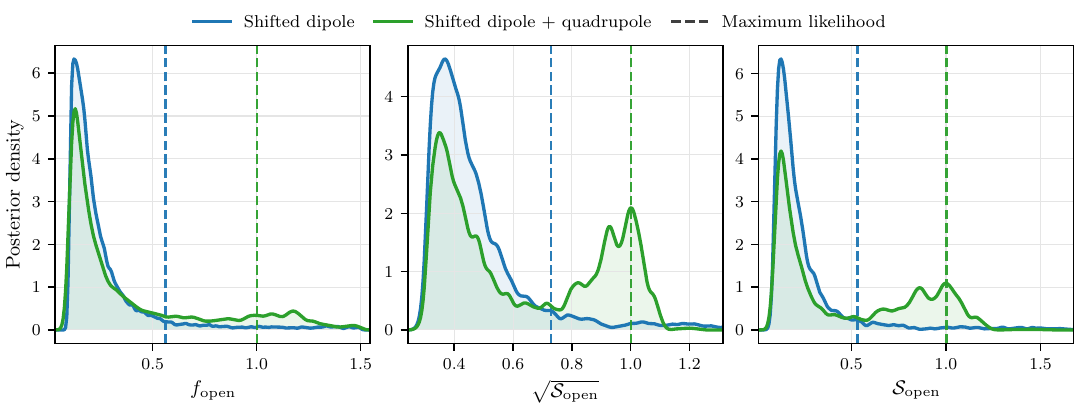}
\caption{Open-flux multiplier, equivalent-area radius ratio, and open-footprint area ratio for the shifted dipole and shifted dipole--quadrupole extension. The extension admits a maximum-likelihood point near the nominal footprint, but its marginalized posterior retains substantial weight at contracted footprints. The maximum-likelihood vectors are listed in Table~\ref{tab:maxlike_parameters}.\label{fig:appendix_footprint_comparison}}
\end{figure*}

\begin{figure*}
\centering
\includegraphics[width=0.98\textwidth]{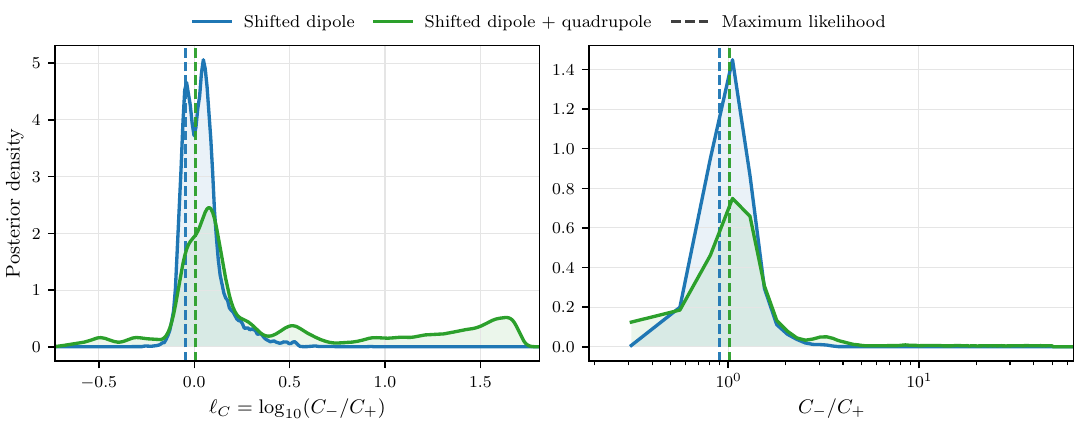}
\caption{Negative-to-positive-$\cos\theta'$ branch temperature-normalization ratios for the shifted dipole and shifted dipole--quadrupole extension. Both have a symmetric mode; the extension also contains a broad high-contrast branch.\label{fig:appendix_cap_ratio_comparison}}
\end{figure*}

Figures~\ref{fig:appendix_mass_radius_comparison}--\ref{fig:appendix_cap_ratio_comparison} compare the extension directly with the shifted dipole. They show that the principal stellar solution and symmetric-heating mode remain stable, while the added quadrupolar freedom opens a high-likelihood path toward the nominal footprint and also produces broader marginalized tails. This is why the extension is physically interesting but not strongly required by the current observations.

\bibliographystyle{aasjournal}
\bibliography{main}

@ARTICLE{ChenBeloborodov2013,
       author = {{Chen}, Alexander Y. and {Beloborodov}, Andrei M.},
        title = "{Dead Zone in the Polar-cap Accelerator of Pulsars}",
      journal = {\apj},
         year = 2013,
        month = jan,
       volume = {762},
       number = {2},
          eid = {76},
        pages = {76},
          doi = {10.1088/0004-637X/762/2/76},
archivePrefix = {arXiv},
       eprint = {1206.6344},
 primaryClass = {astro-ph.HE},
       adsurl = {https://ui.adsabs.harvard.edu/abs/2013ApJ...762...76C}
}

@ARTICLE{Hoogkamer2025,
       author = {{Hoogkamer}, Mariska and {Kini}, Yves and {Salmi}, Tuomo and {Watts}, Anna L. and {Buchner}, Johannes},
        title = "{Cross-comparison of sampling algorithms for pulse profile modeling of PSR J0740+6620}",
      journal = {\prd},
         year = 2025,
        month = jul,
       volume = {112},
       number = {2},
          eid = {023008},
        pages = {023008},
          doi = {10.1103/cp8c-2nbk},
archivePrefix = {arXiv},
       eprint = {2502.13682},
 primaryClass = {astro-ph.HE},
       adsurl = {https://ui.adsabs.harvard.edu/abs/2025PhRvD.112b3008H}
}

@ARTICLE{Gorski2005,
       author = {{G{\'o}rski}, K.~M. and {Hivon}, E. and {Banday}, A.~J. and {Wandelt}, B.~D. and {Hansen}, F.~K. and {Reinecke}, M. and {Bartelmann}, M.},
        title = "{HEALPix: A Framework for High-Resolution Discretization and Fast Analysis of Data Distributed on the Sphere}",
      journal = {\apj},
         year = 2005,
        month = apr,
       volume = {622},
       number = {2},
        pages = {759-771},
          doi = {10.1086/427976},
archivePrefix = {arXiv},
       eprint = {astro-ph/0409513},
 primaryClass = {astro-ph},
       adsurl = {https://ui.adsabs.harvard.edu/abs/2005ApJ...622..759G}
}

@ARTICLE{Pechenick_1987,
       author = {{Pechenick}, K.~R. and {Ftaclas}, C. and {Cohen}, J.~M.},
        title = "{Hot spots on neutron stars - The near-field gravitational lens}",
      journal = {\apj},
         year = 1983,
        month = nov,
       volume = {274},
        pages = {846-857},
          doi = {10.1086/161498},
       adsurl = {https://ui.adsabs.harvard.edu/abs/1983ApJ...274..846P}
}

@ARTICLE{Beloborodov_2002,
       author = {{Beloborodov}, Andrei M.},
        title = "{Gravitational Bending of Light Near Compact Objects}",
      journal = {\apjl},
         year = 2002,
        month = feb,
       volume = {566},
       number = {2},
        pages = {L85-L88},
          doi = {10.1086/339511},
archivePrefix = {arXiv},
       eprint = {astro-ph/0201117},
 primaryClass = {astro-ph},
       adsurl = {https://ui.adsabs.harvard.edu/abs/2002ApJ...566L..85B}
}

@ARTICLE{Poutanen_2006,
       author = {{Poutanen}, Juri and {Beloborodov}, Andrei M.},
        title = "{Pulse profiles of millisecond pulsars and their Fourier amplitudes}",
      journal = {\mnras},
         year = 2006,
        month = dec,
       volume = {373},
       number = {2},
        pages = {836-844},
          doi = {10.1111/j.1365-2966.2006.11088.x},
archivePrefix = {arXiv},
       eprint = {astro-ph/0608663},
 primaryClass = {astro-ph},
       adsurl = {https://ui.adsabs.harvard.edu/abs/2006MNRAS.373..836P}
}

@ARTICLE{Watts_2016,
       author = {{Watts}, Anna L. and {Andersson}, Nils and {Chakrabarty}, Deepto and {Feroci}, Marco and {Hebeler}, Kai and {Israel}, Gianluca and {Lamb}, Frederick K. and {Miller}, M. Coleman and {Morsink}, Sharon and {{\"O}zel}, Feryal and {Patruno}, Alessandro and {Poutanen}, Juri and {Psaltis}, Dimitrios and {Schwenk}, Achim and {Steiner}, Andrew W. and {Stella}, Luigi and {Tolos}, Laura and {van der Klis}, Michiel},
        title = "{Colloquium: Measuring the neutron star equation of state using x-ray timing}",
      journal = {Reviews of Modern Physics},
         year = 2016,
        month = apr,
       volume = {88},
       number = {2},
          eid = {021001},
        pages = {021001},
          doi = {10.1103/RevModPhys.88.021001},
archivePrefix = {arXiv},
       eprint = {1602.01081},
 primaryClass = {astro-ph.HE},
       adsurl = {https://ui.adsabs.harvard.edu/abs/2016RvMP...88b1001W}
}

@ARTICLE{Bogdanov_21,
       author = {{Bogdanov}, Slavko and {Dittmann}, Alexander J. and {Ho}, Wynn C.~G. and {Lamb}, Frederick K. and {Mahmoodifar}, Simin and {Miller}, M. Coleman and {Morsink}, Sharon M. and {Riley}, Thomas E. and {Strohmayer}, Tod E. and {Watts}, Anna L. and {Choudhury}, Devarshi and {Guillot}, Sebastien and {Harding}, Alice K. and {Ray}, Paul S. and {Wadiasingh}, Zorawar and {Wolff}, Michael T. and {Markwardt}, Craig B. and {Arzoumanian}, Zaven and {Gendreau}, Keith C.},
        title = "{Constraining the Neutron Star Mass-Radius Relation and Dense Matter Equation of State with NICER. III. Model Description and Verification of Parameter Estimation Codes}",
      journal = {\apjl},
         year = 2021,
        month = 6,
       volume = {914},
       number = {1},
          eid = {L15},
        pages = {L15},
          doi = {10.3847/2041-8213/abfb79},
archivePrefix = {arXiv},
       eprint = {2104.06928},
 primaryClass = {astro-ph.HE},
       adsurl = {https://ui.adsabs.harvard.edu/abs/2021ApJ...914L..15B}
}

@ARTICLE{Bogdanov_19,
       author = {{Bogdanov}, Slavko and {Lamb}, Frederick K. and {Mahmoodifar}, Simin and {Miller}, M. Coleman and {Morsink}, Sharon M. and {Riley}, Thomas E. and {Strohmayer}, Tod E. and {Tung}, Albert K. and {Watts}, Anna L. and {Dittmann}, Alexander J. and {Chakrabarty}, Deepto and {Guillot}, Sebastien and {Arzoumanian}, Zaven and {Gendreau}, Keith C.},
        title = "{Constraining the Neutron Star Mass-Radius Relation and Dense Matter Equation of State with NICER. II. Emission from Hot Spots on a Rapidly Rotating Neutron Star}",
      journal = {\apjl},
         year = 2019,
        month = dec,
       volume = {887},
       number = {1},
          eid = {L26},
        pages = {L26},
          doi = {10.3847/2041-8213/ab5968},
archivePrefix = {arXiv},
       eprint = {1912.05707},
 primaryClass = {astro-ph.HE},
       adsurl = {https://ui.adsabs.harvard.edu/abs/2019ApJ...887L..26B}
}

@ARTICLE{wilms_00,
       author = {{Wilms}, J. and {Allen}, A. and {McCray}, R.},
        title = "{On the Absorption of X-Rays in the Interstellar Medium}",
      journal = {\apj},
         year = 2000,
        month = oct,
       volume = {542},
       number = {2},
        pages = {914-924},
          doi = {10.1086/317016},
archivePrefix = {arXiv},
       eprint = {astro-ph/0008425},
 primaryClass = {astro-ph},
       adsurl = {https://ui.adsabs.harvard.edu/abs/2000ApJ...542..914W}
}

@ARTICLE{AlGendy_14,
       author = {{AlGendy}, Mohammad and {Morsink}, Sharon M.},
        title = "{Universality of the Acceleration due to Gravity on the Surface of a Rapidly Rotating Neutron Star}",
      journal = {\apj},
         year = 2014,
        month = aug,
       volume = {791},
       number = {2},
          eid = {78},
        pages = {78},
          doi = {10.1088/0004-637X/791/2/78},
archivePrefix = {arXiv},
       eprint = {1404.0609},
 primaryClass = {astro-ph.HE},
       adsurl = {https://ui.adsabs.harvard.edu/abs/2014ApJ...791...78A}
}

@ARTICLE{Ho_2001,
       author = {{Ho}, Wynn C.~G. and {Lai}, Dong},
        title = "{Atmospheres and spectra of strongly magnetized neutron stars}",
      journal = {\mnras},
         year = 2001,
        month = nov,
       volume = {327},
       number = {4},
        pages = {1081-1096},
          doi = {10.1046/j.1365-8711.2001.04801.x},
archivePrefix = {arXiv},
       eprint = {astro-ph/0104199},
 primaryClass = {astro-ph},
       adsurl = {https://ui.adsabs.harvard.edu/abs/2001MNRAS.327.1081H}
}

@ARTICLE{Morsink_2007,
       author = {{Morsink}, Sharon M. and {Leahy}, Denis A. and {Cadeau}, Coire and {Braga}, John},
        title = "{The Oblate Schwarzschild Approximation for Light Curves of Rapidly Rotating Neutron Stars}",
      journal = {\apj},
         year = 2007,
        month = jul,
       volume = {663},
       number = {2},
        pages = {1244-1251},
          doi = {10.1086/518648},
archivePrefix = {arXiv},
       eprint = {astro-ph/0703123},
 primaryClass = {astro-ph},
       adsurl = {https://ui.adsabs.harvard.edu/abs/2007ApJ...663.1244M}
}

@ARTICLE{Riley2019,
       author = {{Riley}, T.~E. and {Watts}, A.~L. and {Bogdanov}, S. and {Ray}, P.~S. and {Ludlam}, R.~M. and {Guillot}, S. and {Arzoumanian}, Z. and {Baker}, C.~L. and {Bilous}, A.~V. and {Chakrabarty}, D. and {Gendreau}, K.~C. and {Harding}, A.~K. and {Ho}, W.~C.~G. and {Lattimer}, J.~M. and {Morsink}, S.~M. and {Strohmayer}, T.~E.},
        title = "{A NICER View of PSR J0030+0451: Millisecond Pulsar Parameter Estimation}",
      journal = {The Astrophysical Journal Letters},
         year = 2019,
        month = dec,
       volume = {887},
       number = {1},
          eid = {L21},
        pages = {L21},
          doi = {10.3847/2041-8213/ab481c},
archivePrefix = {arXiv},
       eprint = {1912.05702},
 primaryClass = {astro-ph.HE},
       adsurl = {https://ui.adsabs.harvard.edu/abs/2019ApJ...887L..21R}
}

@ARTICLE{Riley2021,
       author = {{Riley}, Thomas E. and {Watts}, Anna L. and {Ray}, Paul S. and {Bogdanov}, Slavko and {Guillot}, Sebastien and {Morsink}, Sharon M. and {Bilous}, Anna V. and {Arzoumanian}, Zaven and {Choudhury}, Devarshi and {Deneva}, Julia S. and {Gendreau}, Keith C. and {Harding}, Alice K. and {Ho}, Wynn C.~G. and {Lattimer}, James M. and {Loewenstein}, Michael and {Ludlam}, Renee M. and {Markwardt}, Craig B. and {Okajima}, Takashi and {Prescod-Weinstein}, Chanda and {Remillard}, Ronald A. and {Wolff}, Michael T. and {Fonseca}, Emmanuel and {Cromartie}, H. Thankful and {Kerr}, Matthew and {Pennucci}, Timothy T. and {Parthasarathy}, Aditya and {Ransom}, Scott and {Stairs}, Ingrid and {Guillemot}, Lucas and {Cognard}, Ismael},
        title = "{A NICER View of the Massive Pulsar PSR J0740+6620 Informed by Radio Timing and XMM-Newton Spectroscopy}",
      journal = {The Astrophysical Journal Letters},
         year = 2021,
        month = sep,
       volume = {918},
       number = {2},
          eid = {L27},
        pages = {L27},
          doi = {10.3847/2041-8213/ac0a81},
archivePrefix = {arXiv},
       eprint = {2105.06980},
 primaryClass = {astro-ph.HE},
       adsurl = {https://ui.adsabs.harvard.edu/abs/2021ApJ...918L..27R}
}

@ARTICLE{Miller2021,
       author = {{Miller}, M.~C. and {Lamb}, F.~K. and {Dittmann}, A.~J. and {Bogdanov}, S. and {Arzoumanian}, Z. and {Gendreau}, K.~C. and {Guillot}, S. and {Ho}, W.~C.~G. and {Lattimer}, J.~M. and {Loewenstein}, M. and {Morsink}, S.~M. and {Ray}, P.~S. and {Wolff}, M.~T. and {Baker}, C.~L. and {Cazeau}, T. and {Manthripragada}, S. and {Markwardt}, C.~B. and {Okajima}, T. and {Pollard}, S. and {Cognard}, I. and {Cromartie}, H.~T. and {Fonseca}, E. and {Guillemot}, L. and {Kerr}, M. and {Parthasarathy}, A. and {Pennucci}, T.~T. and {Ransom}, S. and {Stairs}, I.},
        title = "{The Radius of PSR J0740+6620 from NICER and XMM-Newton Data}",
      journal = {The Astrophysical Journal Letters},
         year = 2021,
        month = sep,
       volume = {918},
       number = {2},
          eid = {L28},
        pages = {L28},
          doi = {10.3847/2041-8213/ac089b},
archivePrefix = {arXiv},
       eprint = {2105.06979},
 primaryClass = {astro-ph.HE},
       adsurl = {https://ui.adsabs.harvard.edu/abs/2021ApJ...918L..28M}
}

@ARTICLE{Dittmann24,
       author = {{Dittmann}, Alexander J. and {Miller}, M. Coleman and {Lamb}, Frederick K. and {Holt}, Isiah M. and {Chirenti}, Cecilia and {Wolff}, Michael T. and {Bogdanov}, Slavko and {Guillot}, Sebastien and {Ho}, Wynn C.~G. and {Morsink}, Sharon M. and {Arzoumanian}, Zaven and {Gendreau}, Keith C.},
        title = "{A More Precise Measurement of the Radius of PSR J0740+6620 Using Updated NICER Data}",
      journal = {\apj},
         year = 2024,
        month = oct,
       volume = {974},
       number = {2},
          eid = {295},
        pages = {295},
          doi = {10.3847/1538-4357/ad5f1e},
archivePrefix = {arXiv},
       eprint = {2406.14467},
 primaryClass = {astro-ph.HE},
       adsurl = {https://ui.adsabs.harvard.edu/abs/2024ApJ...974..295D}
}

@article{Salmi_2024_j0740,
   title={The Radius of the High-mass Pulsar PSR J0740+6620 with 3.6 yr of NICER Data},
   volume={974},
   ISSN={1538-4357},
   url={http://dx.doi.org/10.3847/1538-4357/ad5f1f},
   DOI={10.3847/1538-4357/ad5f1f},
   number={2},
   journal={The Astrophysical Journal},
   publisher={American Astronomical Society},
   author={Salmi, Tuomo and Choudhury, Devarshi and Kini, Yves and Riley, Thomas E. and Vinciguerra, Serena and Watts, Anna L. and Wolff, Michael T. and Arzoumanian, Zaven and Bogdanov, Slavko and Chakrabarty, Deepto and Gendreau, Keith and Guillot, Sebastien and Ho, Wynn C. G. and Huppenkothen, Daniela and Ludlam, Renee M. and Morsink, Sharon M. and Ray, Paul S.},
   year={2024},
   month=oct, pages={294} }

@ARTICLE{Fonseca_21,
       author = {{Fonseca}, E. and {Cromartie}, H.~T. and {Pennucci}, T.~T. and {Ray}, P.~S. and {Kirichenko}, A. Yu. and {Ransom}, S.~M. and {Demorest}, P.~B. and {Stairs}, I.~H. and {Arzoumanian}, Z. and {Guillemot}, L. and {Parthasarathy}, A. and {Kerr}, M. and {Cognard}, I. and {Baker}, P.~T. and {Blumer}, H. and {Brook}, P.~R. and {DeCesar}, M. and {Dolch}, T. and {Dong}, F.~A. and {Ferrara}, E.~C. and {Fiore}, W. and {Garver-Daniels}, N. and {Good}, D.~C. and {Jennings}, R. and {Jones}, M.~L. and {Kaspi}, V.~M. and {Lam}, M.~T. and {Lorimer}, D.~R. and {Luo}, J. and {McEwen}, A. and {McKee}, J.~W. and {McLaughlin}, M.~A. and {McMann}, N. and {Meyers}, B.~W. and {Naidu}, A. and {Ng}, C. and {Nice}, D.~J. and {Pol}, N. and {Radovan}, H.~A. and {Shapiro-Albert}, B. and {Tan}, C.~M. and {Tendulkar}, S.~P. and {Swiggum}, J.~K. and {Wahl}, H.~M. and {Zhu}, W.~W.},
        title = "{Refined Mass and Geometric Measurements of the High-mass PSR J0740+6620}",
      journal = {\apjl},
         year = 2021,
        month = jul,
       volume = {915},
       number = {1},
          eid = {L12},
        pages = {L12},
          doi = {10.3847/2041-8213/ac03b8},
archivePrefix = {arXiv},
       eprint = {2104.00880},
 primaryClass = {astro-ph.HE},
       adsurl = {https://ui.adsabs.harvard.edu/abs/2021ApJ...915L..12F}
}

@article{Chen_2020,
   title={A Numerical Model for the Multiwavelength Lightcurves of PSR J0030+0451},
   volume={893},
   ISSN={2041-8213},
   url={http://dx.doi.org/10.3847/2041-8213/ab85c5},
   DOI={10.3847/2041-8213/ab85c5},
   number={2},
   journal={The Astrophysical Journal Letters},
   publisher={American Astronomical Society},
   author={Chen, Alexander Y. and Yuan, Yajie and Vasilopoulos, Georgios},
   year={2020},
   month=apr, pages={L38} }

@article{Olmschenk:2025eie,
   title={Pioneering High-speed Pulsar Parameter Estimation Using Convolutional Neural Networks},
   volume={991},
   ISSN={1538-4357},
   url={http://dx.doi.org/10.3847/1538-4357/ae03c0},
   DOI={10.3847/1538-4357/ae03c0},
   number={2},
   journal={The Astrophysical Journal},
   publisher={American Astronomical Society},
   author={Olmschenk, Greg and Broadbent, Emily and Kalapotharakos, Constantinos and Wallace, Wendy F. and Lechien, Thibault and Wadiasingh, Zorawar and Kazanas, Demosthenes and Harding, Alice},
   year={2025},
   month=sep, pages={169} }

@ARTICLE{Lockhart2019,
       author = {{Lockhart}, Will and {Gralla}, Samuel E. and {{\"O}zel}, Feryal and {Psaltis}, Dimitrios},
        title = "{X-ray light curves from realistic polar cap models: inclined pulsar magnetospheres and multipole fields}",
      journal = {\mnras},
         year = 2019,
        month = dec,
       volume = {490},
       number = {2},
        pages = {1774-1783},
          doi = {10.1093/mnras/stz2524},
archivePrefix = {arXiv},
       eprint = {1904.11534},
 primaryClass = {astro-ph.HE},
       adsurl = {https://ui.adsabs.harvard.edu/abs/2019MNRAS.490.1774L}
}

@ARTICLE{Gralla2017,
       author = {{Gralla}, Samuel E. and {Lupsasca}, Alexandru and {Philippov}, Alexander},
        title = "{Inclined Pulsar Magnetospheres in General Relativity: Polar Caps for the Dipole, Quadrudipole, and Beyond}",
      journal = {\apj},
         year = 2017,
        month = dec,
       volume = {851},
       number = {2},
          eid = {137},
        pages = {137},
          doi = {10.3847/1538-4357/aa978d},
archivePrefix = {arXiv},
       eprint = {1704.05062},
 primaryClass = {astro-ph.HE},
       adsurl = {https://ui.adsabs.harvard.edu/abs/2017ApJ...851..137G}
}

@misc{UltraNest,
      title={UltraNest -- a robust, general purpose Bayesian inference engine}, 
      author={Johannes Buchner},
      year={2021},
      eprint={2101.09604},
      archivePrefix={arXiv},
      primaryClass={stat.CO},
      url={https://arxiv.org/abs/2101.09604}, 
}

@article{Huang:2025_hotspot,
    author = "Huang, Chun and Chen, Alexander Y.",
    title = "{Physics-motivated Models of Pulsar X-Ray Hotspots: Off-center Dipole Configurations}",
    eprint = "2502.15881",
    archivePrefix = "arXiv",
    primaryClass = "astro-ph.HE",
    doi = "10.3847/1538-4357/adf747",
    journal = "Astrophys. J.",
    volume = "991",
    number = "1",
    pages = "90",
    year = "2025"
}

@article{Kalapotharakos_2021,
   title={The Multipolar Magnetic Field of the Millisecond Pulsar PSR J0030+0451},
   volume={907},
   ISSN={1538-4357},
   url={http://dx.doi.org/10.3847/1538-4357/abcec0},
   DOI={10.3847/1538-4357/abcec0},
   number={2},
   journal={The Astrophysical Journal},
   publisher={American Astronomical Society},
   author={Kalapotharakos, Constantinos and Wadiasingh, Zorawar and Harding, Alice K. and Kazanas, Demosthenes},
   year={2021},
   month=jan, pages={63} }

@ARTICLE{Miller2019,
       author = {{Miller}, M.~C. and {Lamb}, F.~K. and {Dittmann}, A.~J. and {Bogdanov}, S. and {Arzoumanian}, Z. and {Gendreau}, K.~C. and {Guillot}, S. and {Harding}, A.~K. and {Ho}, W.~C.~G. and {Lattimer}, J.~M. and {Ludlam}, R.~M. and {Mahmoodifar}, S. and {Morsink}, S.~M. and {Ray}, P.~S. and {Strohmayer}, T.~E. and {Wood}, K.~S. and {Enoto}, T. and {Foster}, R. and {Okajima}, T. and {Prigozhin}, G. and {Soong}, Y.},
        title = "{PSR J0030+0451 Mass and Radius from NICER Data and Implications for the Properties of Neutron Star Matter}",
      journal = {The Astrophysical Journal Letters},
         year = 2019,
        month = dec,
       volume = {887},
       number = {1},
          eid = {L24},
        pages = {L24},
          doi = {10.3847/2041-8213/ab50c5},
archivePrefix = {arXiv},
       eprint = {1912.05705},
 primaryClass = {astro-ph.HE},
       adsurl = {https://ui.adsabs.harvard.edu/abs/2019ApJ...887L..24M}
}

@ARTICLE{Taiyebah2026,
       author = {{Taiyebah}, Farhana and {Kalapotharakos}, Constantinos and {Olmschenk}, Greg and {Wallace}, Wendy F. and {De}, Soumi and {Bucker Siddik}, Abu and {Oyen}, Diane and {Lechien}, Thibault and {Wadiasingh}, Zorawar},
        title = "{Multipolar Magnetic-Field Inference for PSR J0740+6620 with Neural-Network-Accelerated NICER Pulse-Profile Modeling}",
      journal = {arXiv e-prints},
         year = 2026,
        month = jun,
          eid = {arXiv:2606.30886},
        pages = {arXiv:2606.30886},
          doi = {10.48550/arXiv.2606.30886},
archivePrefix = {arXiv},
       eprint = {2606.30886},
 primaryClass = {astro-ph.HE},
       adsurl = {https://ui.adsabs.harvard.edu/abs/2026arXiv260630886T}
}

@ARTICLE{zhou2026,
       author = {{Zhou}, Tianzhe and {Huang}, Chun},
        title = "{GPU-accelerated X-ray pulse profile modeling}",
      journal = {\aap},
         year = 2026,
        month = may,
       volume = {709},
          eid = {A111},
        pages = {A111},
          doi = {10.1051/0004-6361/202659305},
archivePrefix = {arXiv},
       eprint = {2510.07764},
 primaryClass = {astro-ph.HE},
       adsurl = {https://ui.adsabs.harvard.edu/abs/2026A&A...709A.111Z}
}

@ARTICLE{Huang2026,
       author = {{Huang}, Chun},
        title = "{First-principles Polar-cap Currents in Multipolar Pulsar Magnetospheres}",
      journal = {\apj},
         year = 2026,
        month = mar,
       volume = {999},
       number = {2},
          eid = {204},
        pages = {204},
          doi = {10.3847/1538-4357/ae4489},
archivePrefix = {arXiv},
       eprint = {2602.12376},
 primaryClass = {astro-ph.HE},
       adsurl = {https://ui.adsabs.harvard.edu/abs/2026ApJ...999..204H}
}

@article{Guillemot_2016,
   title={The gamma-ray millisecond pulsar deathline, revisited: New velocity and distance measurements},
   volume={587},
   ISSN={1432-0746},
   url={http://dx.doi.org/10.1051/0004-6361/201527847},
   DOI={10.1051/0004-6361/201527847},
   journal={Astronomy and Astrophysics},
   publisher={EDP Sciences},
   author={Guillemot, L. and Smith, D. A. and Laffon, H. and Janssen, G. H. and Cognard, I. and Theureau, G. and Desvignes, G. and Ferrara, E. C. and Ray, P. S.},
   year={2016},
   month=Feb, pages={A109} }

@article{YeChen2025,
    author = "Ye, Dingyi and Chen, Alexander Y.",
    title = "{1D Vlasov Simulations of QED Cascades over Pulsar Polar Caps}",
    eprint = "2507.15804",
    archivePrefix = "arXiv",
    primaryClass = "astro-ph.HE",
    doi = "10.3847/1538-4357/ae19e5",
    journal = "Astrophys. J.",
    volume = "995",
    number = "2",
    pages = "179",
    year = "2025"
}

@ARTICLE{Ruderman1975,
       author = {{Ruderman}, M.~A. and {Sutherland}, P.~G.},
        title = "{Theory of pulsars: polar gaps, sparks, and coherent microwave radiation.}",
      journal = {\apj},
         year = 1975,
        month = feb,
       volume = {196},
        pages = {51-72},
          doi = {10.1086/153393},
       adsurl = {https://ui.adsabs.harvard.edu/abs/1975ApJ...196...51R}
}

@INPROCEEDINGS{Harding2022,
       author = {{Harding}, Alice K.},
        title = "{The Emission Physics of Millisecond Pulsars}",
    booktitle = {Astrophysics and Space Science Library},
         year = 2022,
       editor = {{Bhattacharyya}, Sudip and {Papitto}, Alessandro and {Bhattacharya}, Dipankar},
       series = {Astrophysics and Space Science Library},
       volume = {465},
        month = jan,
        pages = {57-85},
          doi = {10.1007/978-3-030-85198-9_3},
archivePrefix = {arXiv},
       eprint = {2101.05751},
 primaryClass = {astro-ph.HE},
       adsurl = {https://ui.adsabs.harvard.edu/abs/2022ASSL..465...57H}
}

@ARTICLE{Harding_2002,
       author = {{Harding}, Alice K. and {Muslimov}, Alexander G. and {Zhang}, Bing},
        title = "{Regimes of Pulsar Pair Formation and Particle Energetics}",
      journal = {\apj},
         year = 2002,
        month = sep,
       volume = {576},
       number = {1},
        pages = {366-375},
          doi = {10.1086/341633},
archivePrefix = {arXiv},
       eprint = {astro-ph/0205077},
 primaryClass = {astro-ph},
       adsurl = {https://ui.adsabs.harvard.edu/abs/2002ApJ...576..366H}
}

@article{Erber1966,
  title = {High-Energy Electromagnetic Conversion Processes in Intense Magnetic Fields},
  author = {Erber, THOMAS},
  journal = {Rev. Mod. Phys.},
  volume = {38},
  issue = {4},
  pages = {626--659},
  numpages = {0},
  year = {1966},
  month = {Oct},
  publisher = {American Physical Society},
  doi = {10.1103/RevModPhys.38.626},
  url = {https://link.aps.org/doi/10.1103/RevModPhys.38.626}
}

@ARTICLE{Petri_2025_pulsar,
       author = {{P{\'e}tri}, J. and {Guillot}, S. and {Guillemot}, L. and {Gonz{\'a}lez-Caniulef}, D. and {Jankowski}, F. and {Grie{\ss}meier}, J.-M. and {Theureau}, G. and {Cognard}, I.},
        title = "{A double dipole geometry for PSR J0740+6620}",
      journal = {\aap},
         year = 2025,
        month = sep,
       volume = {701},
          eid = {A39},
        pages = {A39},
          doi = {10.1051/0004-6361/202555574},
archivePrefix = {arXiv},
       eprint = {2507.10197},
 primaryClass = {astro-ph.HE},
       adsurl = {https://ui.adsabs.harvard.edu/abs/2025A&A...701A..39P}
}

@misc{Buchner2021,
      title={UltraNest -- a robust, general purpose Bayesian inference engine}, 
      author={Johannes Buchner},
      year={2021},
      eprint={2101.09604},
      archivePrefix={arXiv},
      primaryClass={stat.CO},
      url={https://arxiv.org/abs/2101.09604}, 
}

@ARTICLE{Gralla2016,
       author = {{Gralla}, Samuel E. and {Lupsasca}, Alexandru and {Philippov}, Alexander},
        title = "{Pulsar Magnetospheres: Beyond the Flat Spacetime Dipole}",
      journal = {\apj},
         year = 2016,
        month = dec,
       volume = {833},
       number = {2},
          eid = {258},
        pages = {258},
          doi = {10.3847/1538-4357/833/2/258},
archivePrefix = {arXiv},
       eprint = {1604.04625},
 primaryClass = {astro-ph.HE},
       adsurl = {https://ui.adsabs.harvard.edu/abs/2016ApJ...833..258G}
}

@article{Fonseca2021,
   title={Refined Mass and Geometric Measurements of the High-mass PSR J0740+6620},
   volume={915},
   ISSN={2041-8213},
   url={http://dx.doi.org/10.3847/2041-8213/ac03b8},
   DOI={10.3847/2041-8213/ac03b8},
   number={1},
   journal={The Astrophysical Journal Letters},
   publisher={American Astronomical Society},
   author={Fonseca, E. and Cromartie, H. T. and Pennucci, T. T. and Ray, P. S. and Kirichenko, A. Yu. and Ransom, S. M. and Demorest, P. B. and Stairs, I. H. and Arzoumanian, Z. and Guillemot, L. and Parthasarathy, A. and Kerr, M. and Cognard, I. and Baker, P. T. and Blumer, H. and Brook, P. R. and DeCesar, M. and Dolch, T. and Dong, F. A. and Ferrara, E. C. and Fiore, W. and Garver-Daniels, N. and Good, D. C. and Jennings, R. and Jones, M. L. and Kaspi, V. M. and Lam, M. T. and Lorimer, D. R. and Luo, J. and McEwen, A. and McKee, J. W. and McLaughlin, M. A. and McMann, N. and Meyers, B. W. and Naidu, A. and Ng, C. and Nice, D. J. and Pol, N. and Radovan, H. A. and Shapiro-Albert, B. and Tan, C. M. and Tendulkar, S. P. and Swiggum, J. K. and Wahl, H. M. and Zhu, W. W.},
   year={2021},
    pages={L12} }

@ARTICLE{Timokhin2013,
       author = {{Timokhin}, A.~N. and {Arons}, J.},
        title = "{Current flow and pair creation at low altitude in rotation-powered pulsars' force-free magnetospheres: space charge limited flow}",
      journal = {\mnras},
         year = 2013,
        month = feb,
       volume = {429},
       number = {1},
        pages = {20-54},
          doi = {10.1093/mnras/sts298},
archivePrefix = {arXiv},
       eprint = {1206.5819},
 primaryClass = {astro-ph.HE},
       adsurl = {https://ui.adsabs.harvard.edu/abs/2013MNRAS.429...20T}
}

@ARTICLE{Salmi2026,
       author = {{Salmi}, Tuomo and {N{\"a}ttil{\"a}}, Joonas},
        title = "{Pair Discharges and Radio Emission from Millisecond-Pulsar and White-Dwarf Magnetospheres}",
      journal = {arXiv e-prints},
         year = 2026,
        month = aug,
          eid = {arXiv:2608.22027},
        pages = {arXiv:2608.22027},
          doi = {10.48550/arXiv.2608.22027},
archivePrefix = {arXiv},
       eprint = {2608.22027},
 primaryClass = {astro-ph.HE},
       adsurl = {https://ui.adsabs.harvard.edu/abs/2026arXiv260822027S}
}

@article{P_tri_2020,
   title={Off-centred force-free neutron star magnetospheres},
   volume={501},
   ISSN={1365-2966},
   url={http://dx.doi.org/10.1093/mnras/staa3909},
   DOI={10.1093/mnras/staa3909},
   number={3},
   journal={Monthly Notices of the Royal Astronomical Society},
   publisher={Oxford University Press (OUP)},
   author={Pétri, J},
   year={2020},
   month=Dec, pages={4479–4489} }

\end{document}